\documentclass{aa}  
\usepackage{natbib}
\usepackage{graphicx}
\usepackage{txfonts}
\usepackage{caption}
\usepackage{subcaption}
\usepackage{amsmath,amsfonts,bm}
\usepackage{comment}
 \newcommand{\AU}{au}
 
\usepackage[dvipsnames]{xcolor}   

\begin{document}

   \title{Multi spacecraft study with the Icarus model}

   \subtitle{Modelling the propagation of CMEs to Mercury and Earth}

   \author{T. Baratashvili \inst{1}, B. Grison \inst{2},  B. Schmieder \inst{1,3,4}, P. Démoulin \inst{3}, S. Poedts \inst{1,5}
          }

   \institute{Department of Mathematics/Centre for mathematical Plasma Astrophysics, 
             KU Leuven, 3001 Leuven, Belgium\\
             \email{tinatin.baratashvili@kuleuven.be}
             \and
             Institute of Atmospheric Physics CAS, Dept of Space Physics, 14100 Prague, Czech Republic
             \and
             LESIA, Observatoire de Paris, Universit\'e PSL, CNRS, Sorbonne Universit\'e, Univ. Paris Diderot, Sorbonne Paris Cit\'e, 5 place Jules Janssen, 92195 Meudon, France
             \and
             SUPA, School of Physics and Astronomy, University of Glasgow, G12 8QQ, UK
             \and
             Institute of Physics, University of Maria Curie-Sk{\l}odowska, 
             PL-20 031 Lublin, Poland
             }

   \date{Received: \today}

  \abstract
   {Coronal Mass Ejections (CMEs) are the main drivers of the disturbances in interplanetary space. The Earth-directed CMEs can be dangerous, and understanding the CME interior magnetic structure is crucial for advancing space weather studies. Assessing the capabilities of a numerical heliospheric model is crucial, as understanding the nature and extent of its limitations can be used for improving the model and the space weather predictions based on it. 
   }
   {The present paper aims to test the capabilities of the recently developed heliospheric model Icarus and the linear force-free spheromak model that has been implemented in it. 
   }
   {To validate the Icarus space weather modeling tool, 
   two CME events were selected that were observed by two spacecraft located near Mercury and Earth, respectively. 
   This enables testing the heliospheric model computed with Icarus at two distant locations. The source regions for the CMEs were identified, and the CME parameters 
   were determined and later optimized. Different adaptive mesh refinement levels were applied in the simulations to assess its performance by comparing the simulation results to in-situ measurements. 
   }
   {The first CME event erupted on SOL2013-07-09T15:24. 
   The modeled time series were in good agreement with the observations both at MESSENGER and ACE. 
   The second CME event started on SOL2014-02-16T10:24 and 
   was more complicated, as three CME interactions occurred in this event. 
   It was impossible to recover the observed profiles without modeling the other two CMEs that were observed, one before the main CME and one afterward. The parameters for the three CMEs were identified and the three CMEs were modelled in Icarus. For both CME studies, AMR level~3 was sufficient to reconstruct small-scale features near Mercury, while at Earth, AMR level~4 was necessary due to the radially stretched grid that was used. }
   {The profiles obtained at both spacecraft resemble the in-situ measurements well. 
   The space weather modeling tool's current limitation resulted in a too-small deceleration of the CME propagation during the CME-CME interaction between MESSENGER and ACE.
   }

\maketitle
%




\section{Introduction}

Coronal Mass Ejections (CMEs) are massive eruptive events originating in the solar corona and travelling out into the heliosphere. They are considered the main drivers of space weather disturbances \citep{Gopalswamy2017}. Space weather is the branch of physics that focuses on studying and predicting the physical conditions in the heliosphere, with a particular focus on Earth. Thus, studying the main sources of the disturbances is crucial for advancing space weather studies. CMEs are often associated with solar flares, which are energy bursts within the solar corona. During a CME eruption, a large plasma cloud (up to $10^{16}\;$g) is expelled from the Sun \citep{Webb2012}. Their speeds range between $100 - 3,000\;$km s$^{-1}$ based on SOHO/LASCO observations. The average speed of CMEs was estimated by \citet{Webb2006} to be $\sim  450\;$km s$^{-1}$. CME eruptions occur more frequently during high solar activity periods. During solar minimum, the number of eruptions decreases and the interplanetary medium is less disturbed. 

As a CME travels through space, it interacts with the solar wind. During periods when the Sun is most active, multiple CMEs can be released from nearby regions. When these CMEs come into contact with each other along their propagation in the heliosphere, they create intricate CME-CME interaction regions that travel through the solar wind. CMEs expand as they travel from the Sun, and they can reach the sizes of $\sim 10-20\;$R$_\odot$ already at 0.1~au from the Sun. The exact eruption scenarios and the interior description of the CMEs are still under debate among solar physicists. Still, it is commonly agreed that the violent eruptions are connected with the release of magnetic energy. The CME interior is not homogeneous and is often associated with complex magnetic flux-ropes \citep{Webb2012, Cane2003, Vourlidas2013}. 

The CMEs can be observed with coronagraphs such as the SOHO/LASCO-C2 and C3 coronagraphs \citep{Brueckner1995}, and the coronagraphs mounted on the two satellites of the Solar Terrestrial Relations Observatory \citep[STEREO,][]{Kaiser2005}). 
Later, the interplanetary CMEs (ICMEs) can also be identified from in-situ measurements. They have significant features that can be distinguished from the solar wind features \citep{Bourlaga1991,Zhang2021}. 

There are several limitations to the observation capabilities today. The coronagraph images are useful in estimating or calculating the early evolution characteristics of CMEs. They can provide information about the eruption time, direction, size (spread angle), and speed of the CME. However, these images do not help determine the ejected material's magnetic field configuration or strength. On the other hand, in-situ measurements provide local data on plasma and magnetic field quantities at a specific spacecraft location and time. Therefore, they represent only a trajectory through the enormous magnetic clouds as they pass by the spacecraft, which limits its sampling significantly, as the interior is not homogeneous and evolving. Therefore, modeling the global evolution of a magnetic cloud is a challenge for the solar physics community. 

One way to decrease the uncertainty and increase the number of sampling points is to consider multi-point observations \citep{Bourlaga1982}. Numerous new multi-point missions are being proposed and designed, providing greater insight into future events. Multiple spacecraft can help identify historical events when the same CMEs were observed through their alignment at different locations. This approach has enabled scientists to study the interior of CMEs in greater detail, providing a better sampling of these historical events. Often, the knowledge extracted from these multi-point observations is the motivation for proposing such missions to advance our understanding of the CMEs better. Today, it is rare for such ``fortunate'' alignments to occur, which limits the events that can be investigated through this method. However, several studies describe such events observed by multiple spacecraft \citep{winslow2015,Grison2018,Davies2020,Palmerio2021,Salman2020}. 

CMEs are a global threat to the operation of satellites and ground-based technology on Earth. The CME magnetic field interacts with Earth's magnetosphere, causing geomagnetic storms that can lead to power system failures, telecommunication, and power grid blackouts \citep{Kilpua2017}. Nowadays, with the increased amount of spacecraft and satellites in space, the focus of space weather is not only on Earth, but it shifts towards the whole heliosphere, as multiple points are being sampled simultaneously. The violent events have shortened the operational lifespan of satellites, for example, the geophysical satellite GOES suffered by shortening its operational lifespan by three years. Space weather events can cause malfunction of the transmission systems, like, the PRARE instrument on the European Space Agency's ERS-1 satellite, 
which leads to a permanent failure of determining the position of the satellite. This made the interpretation of the measurements of the instrument rather challenging. 

The damage caused by regular events, especially during phases of maximum activity of the Sun (explained in the next section), has been estimated to accumulate to an economic loss of 10 billion euros per year \citep{press_reference}. The potential loss from future strong Earth-directed events is increasing continuously since our dependence on telecommunications, navigation, and electronic systems is increasing day to day. In 2019, the National Threat and Hazard Identification and Risk Assessment (THIRA) of the US Federal Emergency Management Agency (FEMA) identified space weather as one of the two threats that could potentially disturb our society globally, the other one being a pandemic \citep{THIRA2019}. Therefore, an unexpected major solar event can cause damage locally on Earth but also interfere with ongoing space missions. 

Potential damage can be mitigated if the eruption and its propagation are predicted well in advance, allowing preventive measures. Space weather forecasting has become the standard approach for monitoring the environment surrounding Earth and predicting the arrival time and strength of CMEs. 

Recently a new heliospheric modeling tool, Icarus \citep{Verbeke2022,Baratashvili2022}, was developed at the Centre for Mathematical Plasma-Astrophysics (CmPA, KU Leuven) within the European Heliospheric FORecasting Information Asset project \citep[EUHFORIA,][]{Pomoell2018} as an alternative heliospheric wind and CME evolution model. Icarus is implemented in the MPI-AMRVAC framework \citep{Xia2018} and performs solar wind and CME simulations with advanced techniques, such as grid stretching and adaptive mesh refinement (AMR), to obtain fast predictions at Earth \citep{Baratashvili2022}. A magnetized CME model was integrated into Icarus, which allows the modeling of the CME interior magnetic field and the investigation of its evolution as the CME travels from the Sun towards the Earth \citep{Baratashvili2024}. 

To validate the new heliospheric model Icarus and the implemented magnetized spheromak CME model, two CMEs were chosen that were observed by multiple spacecraft from the catalog presented in \cite{winslow2015}. The first CME case occurred on July 9, 2013, and was observed by the MErcury Surface, Space ENvironment, GEochemistry, and Ranging (MESSENGER) spacecraft near Mercury and the Advanced Composition Explorer (ACE) satellite near Earth. The second event occurred on February 16, 2014, and was observed by MESSENGER and ACE. However, in the latter case, the three-CME interaction was identified near Earth, while at MESSENGER the contribution of the three CME was still isolated. Both CME events are modeled in Icarus, and the time series at MESSENGER and ACE are compared with the observational data. Different AMR levels are applied to estimate its performance at MESSENGER and ACE, considering the large radial separation. 

The paper is organized as follows: Section~\ref{icarus_setup} presents the numerical setup in Icarus as well as the available CME models and advanced techniques for efficient simulations, such as grid stretching and solution adaptive mesh refinement. Section~\ref{2013_july} presents the identification of the source, the obtained parameters for modelling the CMEs, and the final results at MESSENGER and ACE for the first CME event.
The analysis of the second CME event is given in Section~\ref{2014_february}. The conclusions and outlooks following this study are given in Section~\ref{conclusions}. 

\section{Numerical setup in Icarus} \label{icarus_setup}

Icarus \citep{Verbeke2022} is a heliospheric modeling tool implemented in the framework of MPI-AMRVAC \citep{Xia2018}. MPI-AMRVAC, a numerical architecture developed in Fortran, solves partial differential equations on different types of grids with the support of different numerical solvers and flux limiters. Therefore, MPI-AMRVAC is well-suited for astrophysical applications. The domain of Icarus is the same as the heliospheric domain of EUHFORIA. The inner radial boundary is located at 0.1~au, which represents the usual separation point between the coronal and heliospheric models, as beyond this point, the solar wind is super-Alfv\'enic, and all the information travels radially outwards. The domain in the radial direction extends up to 2~au, including the orbit of Mars. The whole $360^\circ$ degrees are spanned in the longitudinal direction, but the area in the latitudinal direction is limited to [$-60^\circ +60^\circ$] degrees. Hence, the poles are excluded, which avoids narrow grid cells (in spherical coordinates) near the poles. \cite{Baratashvili2022sungeo} performed a study of different combinations of numerical schemes and flux limiters to obtain the optimal blending regarding optimal capturing of the shocks with the most efficient performance. As a result, the default setting is fixed to the Total Variation Diminishing Lax-Friedrichs numerical method \citep[TVDLF,][]{Toth1996}, a second-order scheme in time and space, combined with the second-order `woodward' flux limiter \citep{vanleer1977}. 

To initiate the solar wind in the heliosphere, the characteristic plasma description is necessary at the inner boundary. Currently, Icarus uses boundary conditions provided with the Wang-Sheeley-Arge (WSA) semi-empirical coronal model \citep{wang1990,arge2003}. Thus obtained plasma conditions at $0.1\;$au are radially extended to $2\;$au and set as initial conditions for the relaxation phase in which the Magnetohydrodynamics (MHD) equations are solved in the heliospheric domain. In contrast to the original EUHFORIA version, the coordinate frame is chosen to be co-rotating with the Sun, implying that once the relaxation phase finishes, the background wind solution is stationary. The relaxation phase is only the first phase when performing the simulations and it usually lasts $\sim 10 - 14\;$days, viz.\ the time that is necessary for the slow wind stream to traverse the entire domain. The next phase is the CME insertion phase, i.e., 5 days before the actual forecasting phase, any CMEs that occurred in that time window are injected at the inner heliospheric boundary. The final phase of the simulations is the space weather forecasting phase, which starts with the injection of the CME(s) for which the evolution and arrival one wants to predict. 

Considering the many similarities between the original EUHFORIA and the newly implemented Icarus models, it is important to stress the main advantages of the Icarus framework. The co-rotating grid and the use of the magnetic field components as primitive variables simplify the injection of the CMEs. Moreover, MPI-AMRVAC supports advanced numerical techniques to optimize the computational mesh, making the simulations more efficient and accurate. Radial grid stretching and adaptive mesh refinement (AMR) are applied in Icarus. Radial grid stretching avoids cell deformation near the inner and outer boundaries when the domain is radially extended as it maintains a constant aspect ratio between the cell widths and lengths throughout the domain. 

\begin{figure}[ht!]
    \centering
    \includegraphics[width=0.5\textwidth]{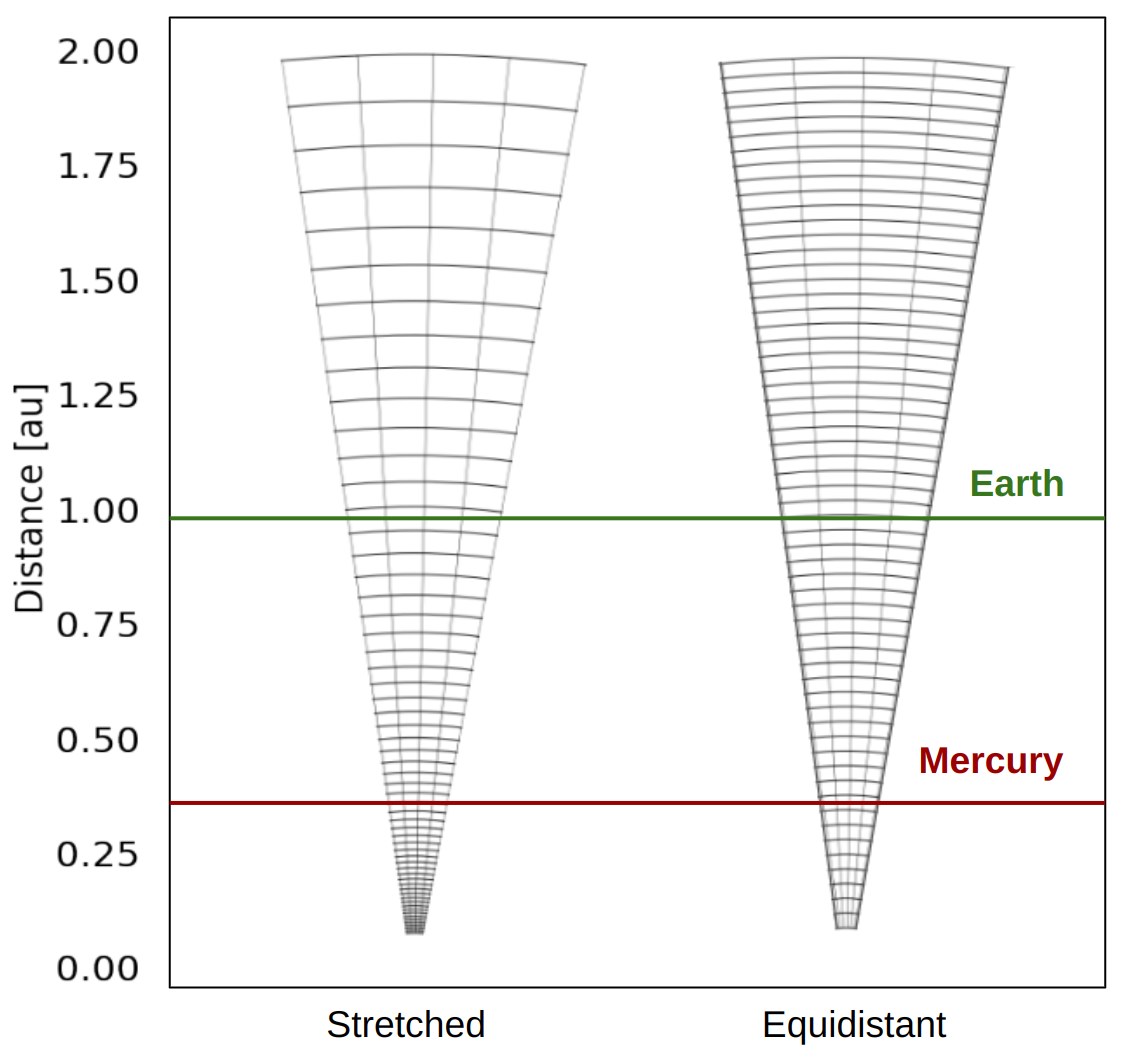}
    \caption{Radial slices cut out from the radially stretched (left) and equidistant (right) grids. The vertical axis shows the distance from the Sun. The type of grid is indicated on the horizontal axis. The red horizontal line shows the location of Mercury along the radius and the green line shows the location of Earth.}
    \label{fig:stretching_mercury_earth}
\end{figure}

Figure~\ref{fig:stretching_mercury_earth} shows cut-out slices from the computational grid for the radially stretched (left) and the equidistant (right) grids. The vertical axis shows the distance from the Sun, and since the heliospheric domain starts at 0.1~au, the first cell starts at 0.1. The red and green horizontal lines show the locations of Mercury and Earth, respectively. In the equidistant slice, the cell sizes are the same in the radial direction at both locations. However, in the radially stretched grid, the difference between the cell sizes in the radial direction at these two locations is notable. As a matter of fact, in the equidistant simulation, the radial width of the cell in the low-resolution simulation is fixed to $1.37\;$R$_\odot$, both at Mercury and Earth, while for the radially stretched grid in the lowest resolution, the radial size of the cell at Mercury is $\sim 4.1\;$R$_\odot$ (depending on the location of Mercury) while at Earth it is $10.7\;$R$_\odot$. This difference is significant for interpreting the numerical simulation results obtained at Mercury and Earth. Compared to the equidistant grid resolution, the basic stretched grid resolution is significantly lower at both locations. On the one hand, this makes the simulations less computationally expensive, but on the other hand, it would decrease the accuracy of the results significantly. 

To compensate for the decreased resolution with the stretched grid, AMR techniques are applied. MPI-AMRVAC is block-adaptive, which means that the AMR is applied using blocks of cells with a maximum of one level of refinement difference between adjacent blocks. MPI-AMRVAC allows the user to define the size of the blocks and to implement the desired criterion. Wherever this criterion is satisfied in the domain, the corresponding area is refined, if necessary up to the indicated maximum refinement level. Different refinement criteria aimed at optimizing CME propagation simulations are discussed in \cite{Baratashvili2022}. Each consecutive AMR level decreases the cell size locally in each direction by a factor of two. Hence, the cell size in any \textit{(n)} AMR level is $(2^{(n-1)})$ times smaller than the local cell size in the basic grid, which is always the low-resolution computational grid. An overview of the used cell sizes in the radial direction at Mercury (averaged at 0.387~au) and Earth (i.e., at L1) for different numbers of AMR levels are given in Table~\ref{table:amr_resolutions_at_earth_messenger}.

\begin{table}[t!]
\caption{Resolutions at L1 for different simulations with the stretched grid. No AMR corresponds to the low-resolution simulation.}   
\label{table:amr_resolutions_at_earth_messenger}   
\centering            
\begin{tabular}{c c c c c}         
\hline
&No AMR & AMR 2 & AMR 3 & AMR 4  \\ 
\hline\hline
  Messenger & 4.14 R$_\odot$ & 2.07R$_\odot$ & 1.03 R$_\odot$ & 0.51R$_\odot$  \\
   Earth &10.7 R$_\odot$ & 5.36R$_\odot$ & 2.68 R$_\odot$ & 1.344R$_\odot$  \\
\hline                                  
\end{tabular}
\end{table}

The radial resolution is significantly lower at Earth than at Mercury, such that one level less refinement at Mercury has a similar resolution to the one level higher refinement resolution at Earth. For comparison, in the low-resolution equidistant mesh simulation, the radial cell size is $1.37\;$R$_\odot$ as stated before, and in the medium-resolution simulation, which is used for standard operational simulations, it is $0.685\;$R$_\odot$ at both locations. 

When AMR is applied optimally, i.e., only at the locations where the higher resolution is necessary, a significant amount of CPU time can be saved \citep{Baratashvili2022,Baratashvili2024}. As the aim of this study is to validate the newly implemented magnetized linear force-free spheromak model in Icarus at multiple locations, the refinement criterion should be tailored to the CME, including both its shock front and interior. Therefore, the combined criterion described in \citet{Baratashvili2022} is applied. Hence, the shock is captured by refining the grid where $\nabla \cdot \vec{V} < 0$, i.e.\ in the compression region. The CME interior is traced with a built-in density tracer function in MPI-AMRVAC. This function traces the injected CME plasma as the CME evolves in the domain and registers the CME plasma density values at these locations, while it remains zero everywhere else. The combination of these two criteria thus follows the whole CME structure (magnetic cloud and leading shock front) throughout the domain. We further constrained the refinement criteria in the space weather modeling scope and restricted the refinement to the Sun-Earth line with a range of $\Delta \phi = 30^\circ$ in the longitudinal direction, $\Delta \theta = 15^\circ$ in the latitudinal direction, and [$50\; R_\odot < r < 250\; R_\odot$] radially. These constraints restrict the domain where the AMR is applied, avoiding unnecessary refinement (spilling CPU time) in the areas that are not interesting to us. 

The magnetized CME model implemented in Icarus is described in \cite{Baratashvili2024}. It is achieved via linking the previously implemented LFF spheromak model in EUHFORIA to Icarus for consistency \citet{Verbeke2019}. Therefore, for obtaining the magnetic field components in Icarus during the injection phase, each time the communication is set up to the EUHFORIA spheromak model, the components are calculated and passed back to Icarus via a Fortran-C binder. The magnetic field is defined as follows in the spheromak model 
\begin{align} \label{B_field_vector}
\mathbf{B} = \frac{1}{r' sin\theta'} \Big [\frac{1}{r'} \frac{\partial \psi}{\partial \theta'} \mathbf{\hat{r}'} - \frac{\partial \psi}{\partial r'} \boldsymbol{\hat{\theta}'} + Q \boldsymbol{\hat{\phi}'} \Big ],
\end{align} 
where $\psi$ and $Q$ are scalar potentials that only depend on $r'$ and $\theta'$ \citep{Chandrasekhar1956}, and $\mathbf{\hat{r}'}$, $\boldsymbol{\hat{\theta}'}$, and $\boldsymbol{\hat{\phi}'}$ denote unit vectors in each of the three local spherical coordinate directions. The magnetic field is divergence-free by the given definition, and the solution is determined so that $\mathbf{J} \, \times \, \mathbf{B} = \mathbf{0}$, thus 'force-free'.

In the simulations presented in this study, the given magnetized CME model is used and the AMR condition is always fixed to the combined criterion described above. 

\section{CME event I: July 9, 2013} \label{2013_july}

\subsection{Observations of the studied CME}

\begin{figure*}
    \centering
    \includegraphics[width=0.33\textwidth]{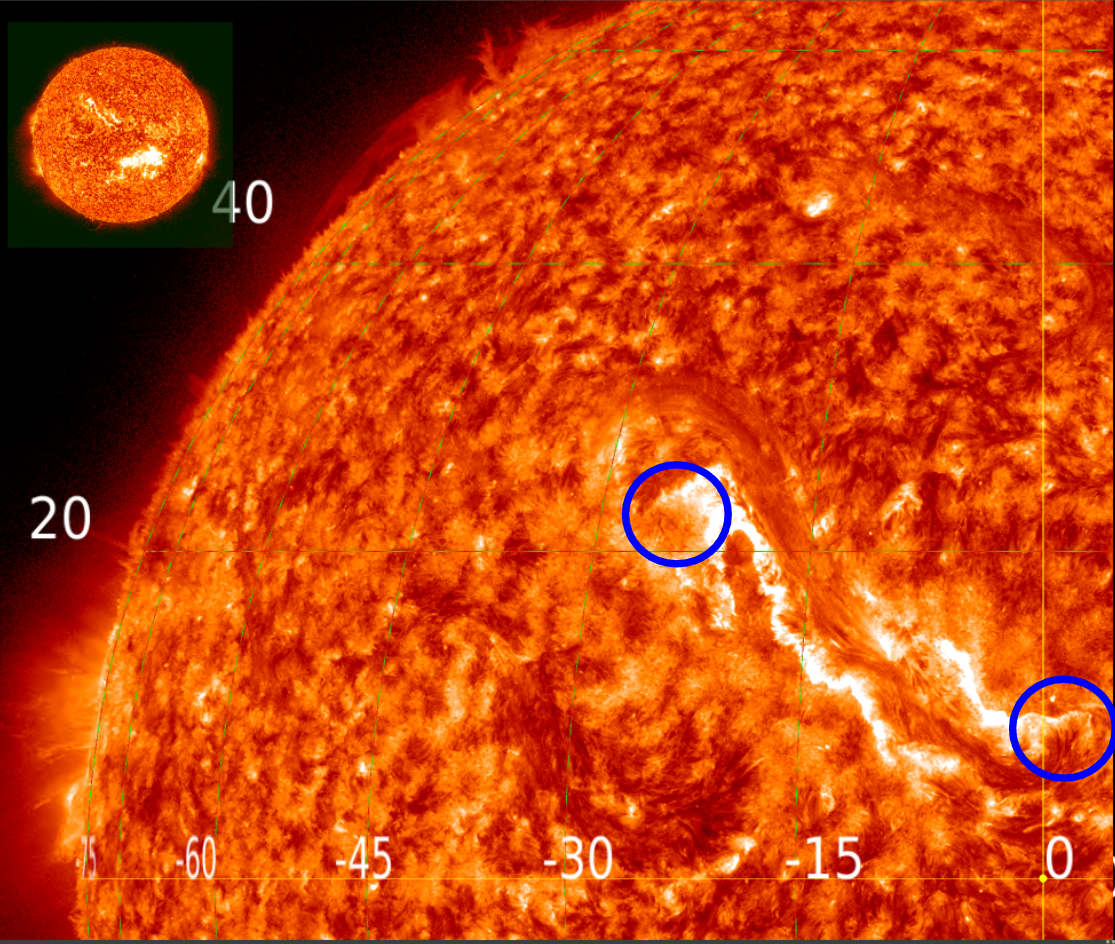}
    \includegraphics[width=0.33\textwidth]{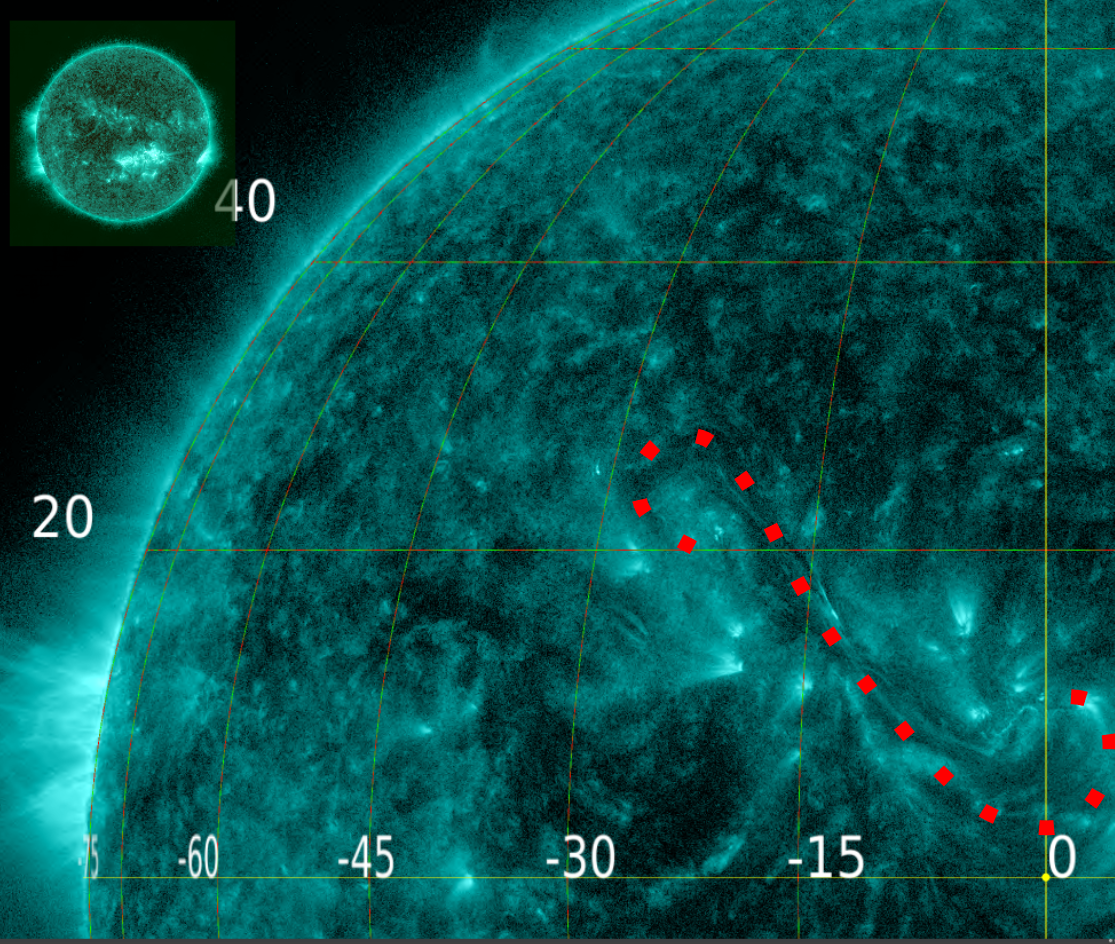}
    \includegraphics[width=0.33\textwidth]{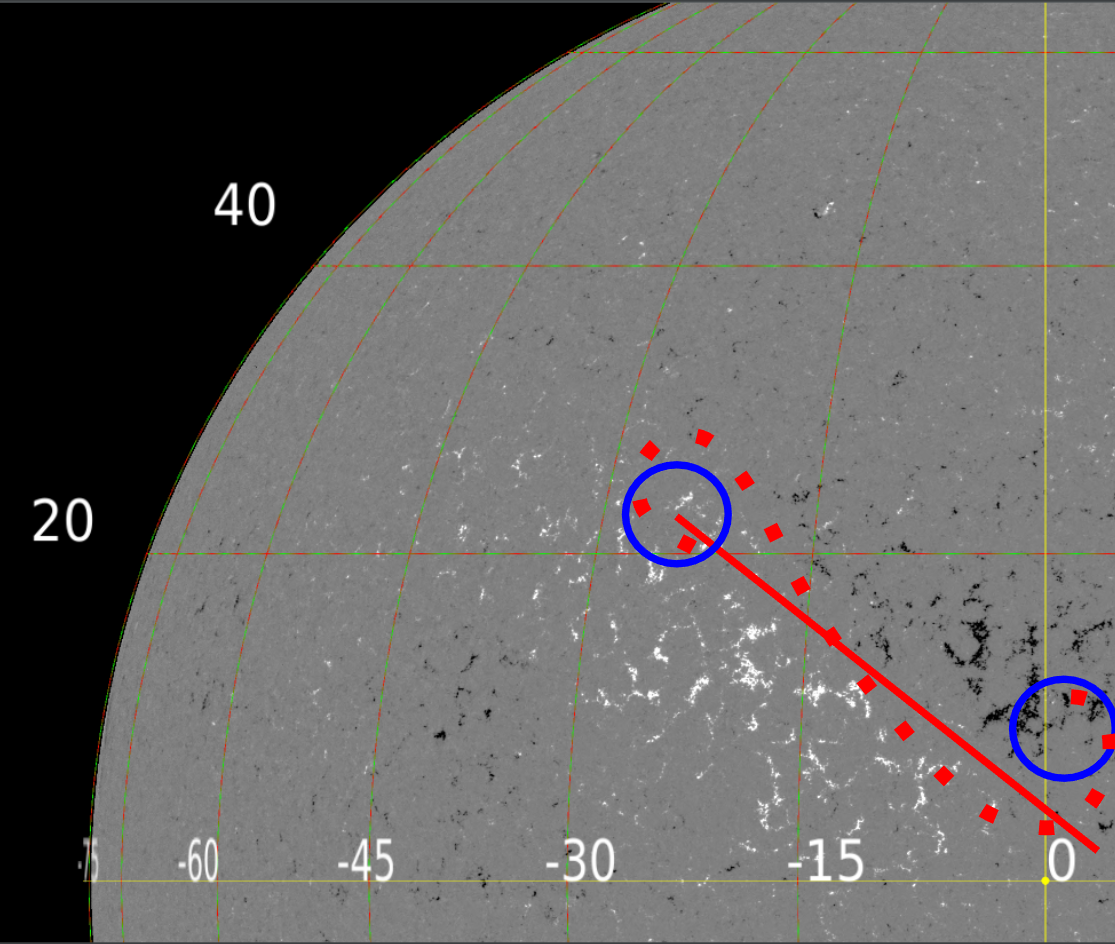}
    \caption{The images taken from JHelioviewer correspond to the source of the CME event prior to the eruption at SOL2013-07-09T13:48:20. The left panel shows the image of the solar atmosphere in Solar Dynamics Observatory/Atmospheric Imaging Assembly 304 \textup{~\AA} wavelength, where the circles identify the footpoints. The middle panel shows the solar corona in SDO/AIA 131 \textup{~\AA} wavelength, with a dotted sigmoid shape plotted on top. The right panel shows the HMI magnetogram with the red line corresponding to the polarity inversion line (PIL) together with the dotted sigmoid and the footprints plotted on the left and middle panels.}
    \label{fig:2013_source}
\end{figure*}
Validating an evolving CME model at different locations in the heliospheric requires a careful choice of the CME event and its extensive analysis. To allow evolution between the observation points, spacecraft with sufficient radial separation were chosen. For the analysis in the present paper, CME events were chosen that have been observed near Mercury by the MESSENGER spacecraft and near the Earth by the ACE spacecraft. 
\cite{Grison2018} found that more than 30 CMEs of the catalog of \cite{winslow2015} have also been observed by VEX, ACE, STEREO-A, or STEREO-B.

For this study, we chose two CME events observed at MESSENGER and ACE, mentioned under CME$\;\#16$ and CME$\;\#19$ in \cite{Grison2018}. 
The first event was observed by MESSENGER starting on July 11, 2013, 01:05 and at ACE starting on July 12, 2013, 16:30 
\citep{Salman2020}. 
The solar wind has been modeled with Icarus using the magnetogram at the time closest to the eruption event. The parameters for the CME have been estimated, and the CME event has been modeled with the Icarus heliospheric modeling tool. 

The associated CME event in the solar corona occurred on July 9, 2013. 
The associated eruption started at 15:24 UT at the Sun according to \citet{winslow2015}. Figure~\ref{fig:2013_source} summarises the observations of the source region. The panels are plotted using data from the Atmospheric Imaging Assembly (AIA) and Helioseismic and Magnetic Imager (HMI) instruments on board the Solar Dynamics Observatory (SDO). The left panel represents the solar atmosphere observed with the 304\textup{~\AA} wavelength filter. The blue circles are plotted at the possible footpoints of the flux rope that erupted. The picture in the middle panel is obtained with the 131\textup{~\AA} filter, where an inverse S-shape sigmoid is visible. The sigmoid was approximated and indicated with the red dotted curve. Using the sigmoid shape of the erupting flux rope is a good proxy to identify the chirality of the CME. As explained by \citet{titov1999} and \citet{demoulin2009}, the inverse S-shape sigmoid can be spotted at the eruption site in the AIA 131 image. This suggests that the sign of the magnetic helicity of the CME is negative and, therefore, the corresponding CME has a left-handed magnetic field orientation. The right panel shows the HMI magnetogram where an approximation of the Polarity Inversion Line (PIL) is over-plotted with a red line to identify the tilt of the eruption \citep{Marubashi2015}. This event is also considered by \citet{Palmerio2018} and the tilt, estimated from the active region configuration at the photosphere, is $-45^\circ$. The inclination of the flux rope is with respect to the ecliptic and the tilt is defined from solar West, opposite to the definition in \citep{Palmerio2018}, where the inclination is defined from solar East and the tilt is identified to be $50^\circ$. 
The approximated sigmoid and the footprint locations are also plotted on top. To constrain the tilt of the spheromak model, we align its magnetic axis with the PIL orientation as done in \cite{Maharana2023} and \cite{Sarkar2024}. For example, a tilt of $0^\circ$ configuration of the spheromak model is a westward flux-rope normal to the Sun–Earth line. For the left-handed flux rope, the inclination is negative (positive) when rotating towards North, anti-clockwise (South, clockwise). 
\begin{figure}[ht!]
    \centering
    \includegraphics[width=0.5\textwidth]{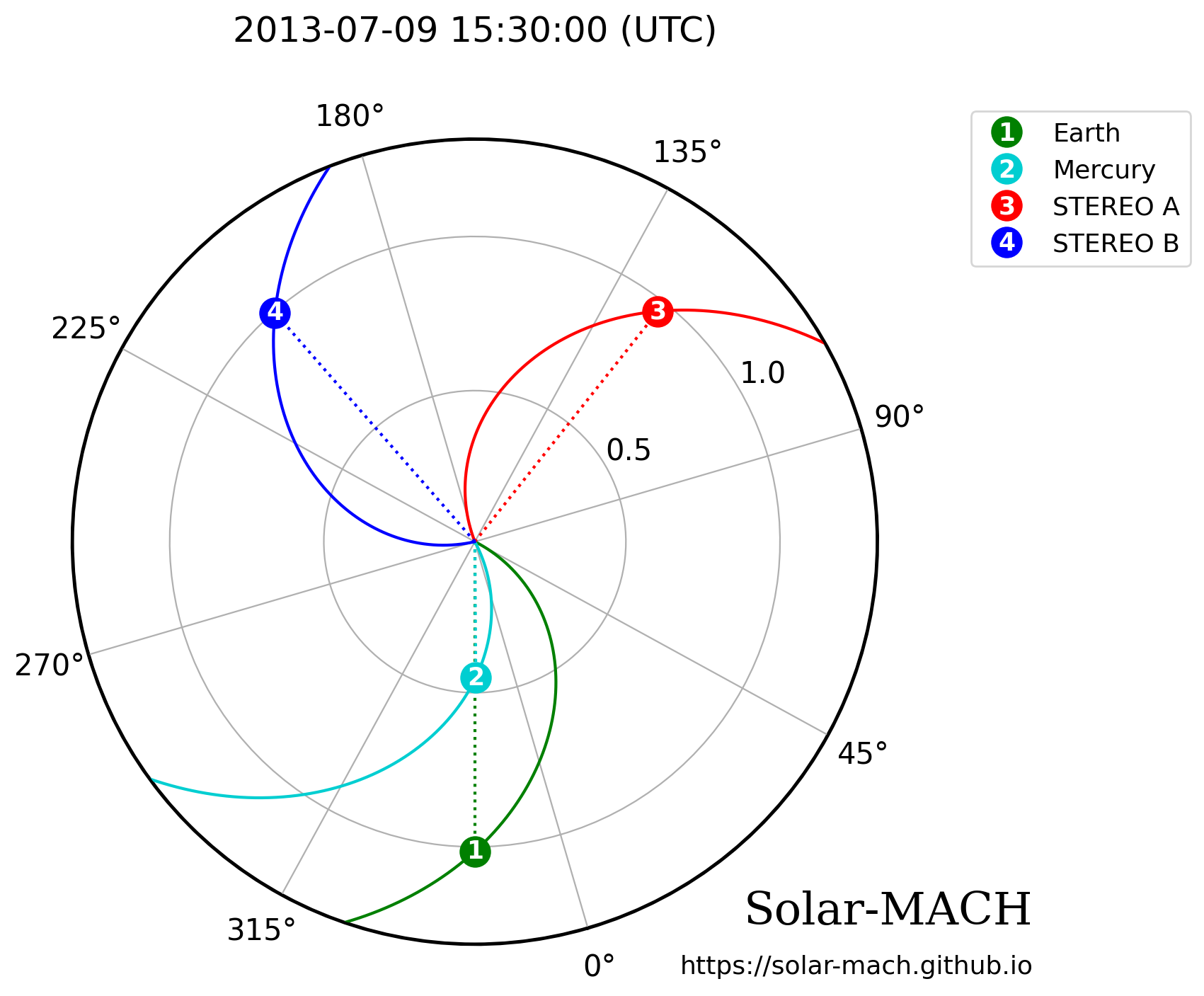}
    \caption{The alignment of the Mercury and Earth at the time of the CME eruption. The locations of STEREO-A and STEREO-B are also plotted. The figure is generated with Solar-MACH \citep{solarmach}.}
    \label{fig:2013_mercuryearth_aligneent}
\end{figure}

The CME event was observed by the STEREO-A and STEREO-B coronagraphs and in situ by MESSENGER near Mercury and ACE near Earth. The locations of STEREO-A, STEREO-B, Mercury, and Earth are given in Figure~\ref{fig:2013_mercuryearth_aligneent}. The angular separation between MESSENGER and ACE at that time is only $3.1^\circ$ \citep{Salman2020}. 

\begin{figure}[ht!]
    \centering
    \includegraphics[width=0.3\textwidth]{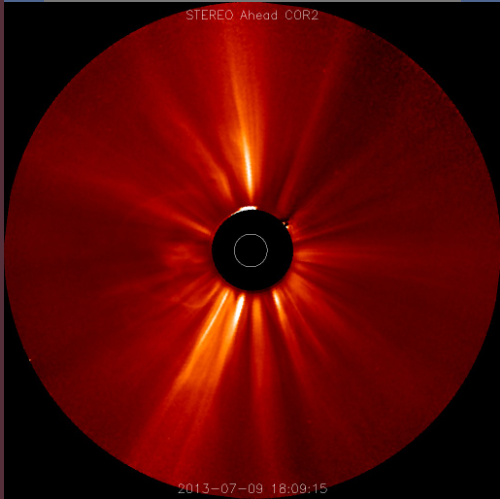}
    \includegraphics[width=0.3\textwidth]{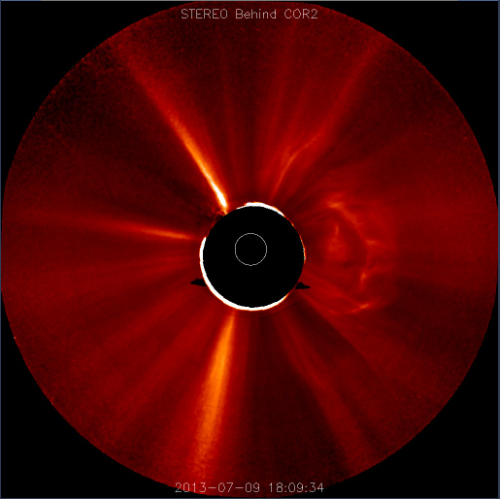}
    \caption{The white-light images seen from STEREO-A C2 and STEREO-B C2 $\sim 3\;$hours after the CME eruption. The figure is generated with the Stereo CME Analysis Tool.}
    \label{fig:2013_sta_stb}
\end{figure}

The images observed by the coronagraph instruments on the STEREO-A and STEREO-B after $\sim 3\;$hours from the CME eruption are plotted in Figure~\ref{fig:2013_sta_stb}. The panels are generated with the Stereo CME Analysis Tool (StereoCAT) at the Community Coordinated Modeling Center.

The CME was fitted with the online tool available on the StereoCAT website to estimate the direction of the CME, its angular width, the speed of propagation, and its arrival time at 0.1~\AU. The initially estimated CME parameters were used as the starting point for the 3D reconstruction with the Graduated Cylindrical Shell (GCS) model \citep{Thernisien2011}. The obtained parameters for injecting the CME at 0.1~au in the heliosphere are given in Table~\ref{table:2013_event_parameters}. As \citet{Palmerio2018} describe, the tilt was estimated from in-situ measurements to be $\sim 10^\circ$ with the axial direction to the west and the flux rope type of NWS, in comparison to the estimated tilt of $50^\circ$ with the axial direction to the southwest and the flux rope type of WSE/NWS from the solar disk appearance. Therefore, we also fixed the tilt to $10^\circ$ in the simulations and later investigated the difference between two models with the same parameters, but the tilt fixed to $45^\circ$ and $10^\circ$ degrees, respectively.

\begin{table}[t!]   
\centering  
\caption{The parameters for the input for the spheromak CME model for the event on July 9, 2013.}   
\begin{tabular}{cc}         
\hline 
Variable & Input value \\ 
\hline  \hline        
    t$_{CME}$ & 2013-07-09T21:21   \\ 
    $\theta_{CME}$ & 15 $^\circ$  \\ 
    $\phi_{CME}$ & 5 $^\circ$  \\
    r$_{CME}$ & 18.0 R$_\odot$ \\ 
    v$_{CME}$ & 300 km s$^{-1}$ \\ 
    $\rho_{CME}$ & 10$^{18}$ kg m$^{-3}$ \\
    T$_{CME}$ & 0.8 $\times 10^6$ K \\ 
    $\tau_{CME}$ & 10$^\circ$ \\ 
    H$_{CME}$ & -1 \\ 
    F$_{CME}$ & $0.5 \times 10^{14}$ Wb\\
\hline                           
\end{tabular}
\label{table:2013_event_parameters}
\end{table}

\subsection{Comparison of simulations to data}

First, the background solar wind was simulated in the heliosphere. The WSA coronal model of EUHFORIA was used to obtain the plasma conditions at 0.1~\AU. The Global Oscillation Network Group (GONG) magnetogram corresponding to 2013-07-09T13:14:00 was used for computing the coronal model. After a relaxation period of 14 days, the spheromak CME model with the parameters described in Table~\ref{table:2013_event_parameters} was injected. The time series modeled at different planets and spacecraft are generated while obtaining the modeled 3D heliosphere. The time series obtained at Mercury and Earth are used to compare with the observed data. 

\begin{figure}
    \centering
    \includegraphics[width=0.5\textwidth]{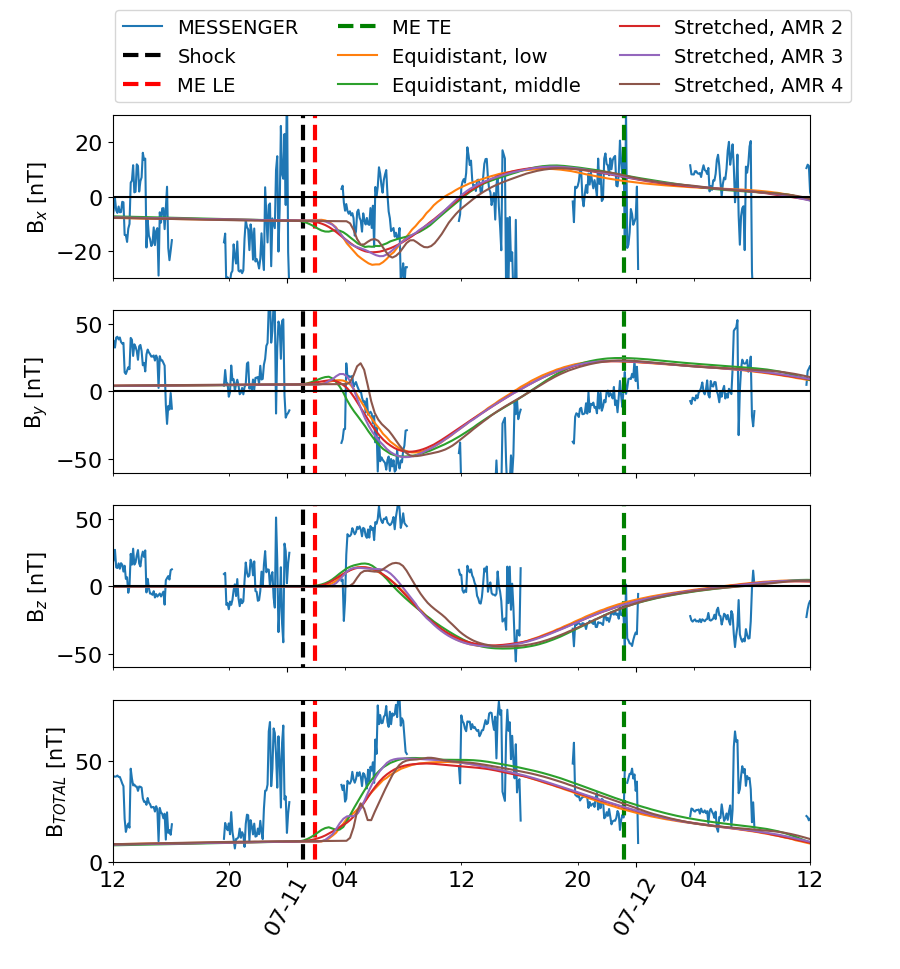}
    \caption{Simulations performed with different numerical grids in Icarus. The blue line corresponds to in-situ data observed at the MESSENGER spacecraft. Magnetic field components are plotted. The black, red, and green lines indicate shock arrival time, the magnetic ejecta leading edge (ME LE), and the magnetic ejecta trailing edge (ME TE), respectively. Equidistant grid simulations on the low and medium-resolution grids are given in orange and green. The simulations performed on the stretched grid with AMR levels 2, 3, and 4 are given in red, purple, and brown, respectively. }
    \label{fig:2013_messenger_b_timeseries}
\end{figure}

Figure~\ref{fig:2013_messenger_b_timeseries} shows the time series of the magnetic field components near Mercury. The blue line corresponds to the observed data by the MESSENGER spacecraft. As MESSENGER was originally designed to study the magnetosphere of Mercury in more detail, it spends a significant amount of time inside the magnetosphere of the planet. The magnetic field of Mercury is rather strong. Therefore, the times when the satellite is affected by the magnetic field of the planet are removed from the observed time series. The nearest magnetopause crossing is estimated to be $\sim1 - 2\;$hours from the spacecraft’s passage through the plasma sheet, during which the plasma parameters and the magnetic field are expected to change \citep{sun2022,james2017}. The average crossing length from the magnetometer observations was estimated to be $\sim 3.5\;$hours. 

The dashed black, red, and green vertical lines indicate the arrival of the disturbance at MESSENGER, followed by the arrival of the Magnetic Ejecta Leading Edge (ME LE) and the passing of the Magnetic Ejecta Trailing Edge (ME TE), according to \cite{winslow2015}. However, the arrival of the shock and the ME LE fall in the period of the magnetosphere passing, which is associated with high spikes in the magnetic field, but does not indicate the arrival of the magnetic cloud. Therefore, the time inside the magnetosphere crossing interval can not be considered as the arrival time of the CME at MESSENGER. The strong variations in the magnetic field components and increased total magnetic field at $\sim$22:00 on July 10, could indicate the arrival of the CME sheath \citep{Bourlaga1982}. 

The results from simulations in Icarus performed on different grids are presented with different curves. The orange and green curves correspond to simulations performed on the low and medium-resolution computational grids, while red, purple, and brown curves correspond to simulations performed on the stretched grids in combination with AMR refinement levels 2, 3, and 4, respectively. The AMR criterion is the same in all simulations, as described in Section~\ref{icarus_setup}. 

Because of the strong noise in the MESSENGER data and the magnetosphere passing at the possible start of the CME arrival at the planet, a detailed comparison of the modeled and observed data is challenging. However, we see that the polarity of the simulated magnetic field for all components is in agreement with the observed data. The strength of the positive $B_z$ component is modeled poorly, as it seems higher in the observed data. However, the negative values of this component are modeled well. The profile of the simulated total magnetic field strength resembles the observed profile. The simulated $B_x$ and $B_y$ components vary rather smoothly in time, and no strong variations are present. The advantage of applied AMR can be seen in the profiles of $B_y$ and the total magnetic field strength at the effective arrival of the magnetic cloud in the simulation. The small peak before the arrival of the magnetic cloud is better resolved in the AMR level 3 simulation than in the equidistant medium-resolution simulation and is more similar to the observational data. The feature in AMR level 4 simulations is sharper than in all other simulation results and shows more variations as MESSENGER passes through the magnetic cloud. 
 
\begin{figure*}
    \centering
    \includegraphics[width=0.48\textwidth]{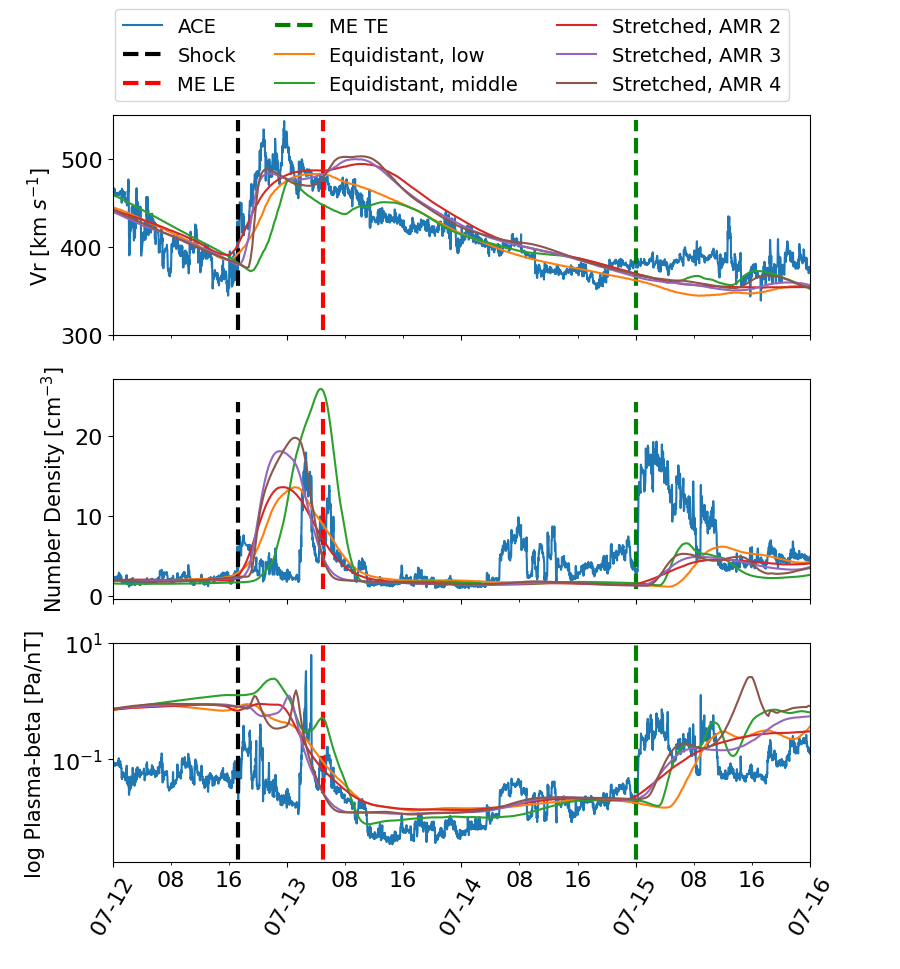}
    \includegraphics[width=0.48\textwidth]{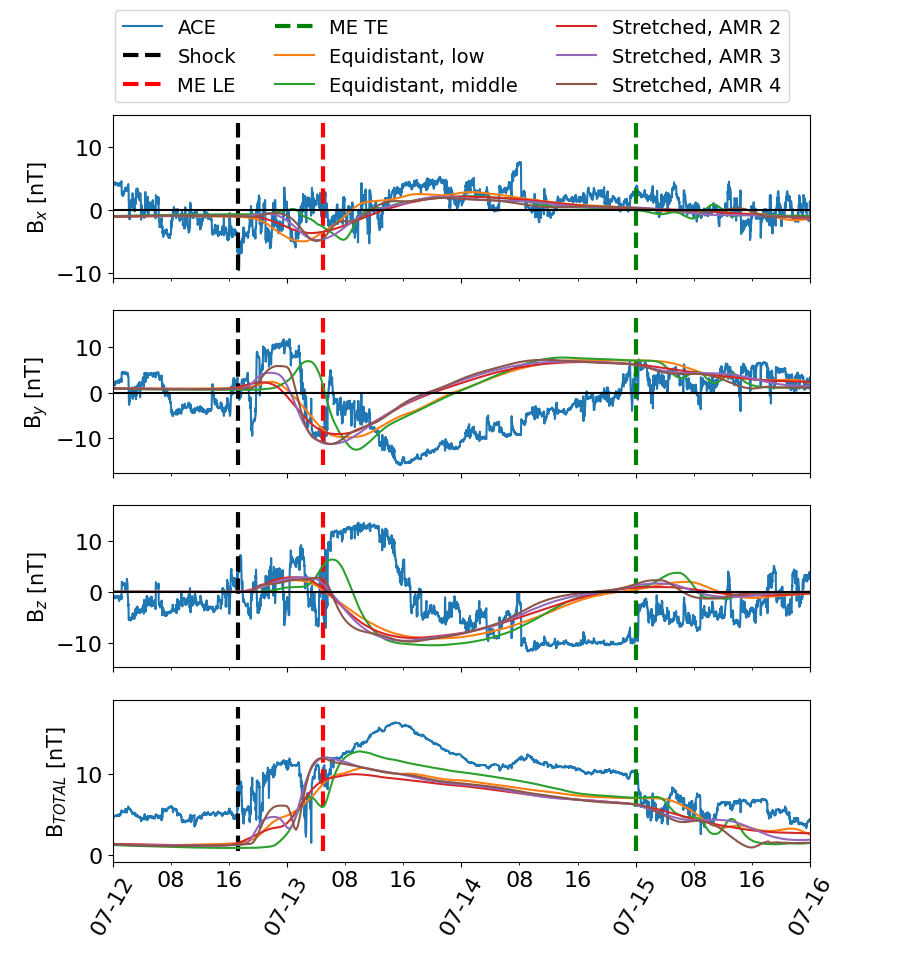}
    \caption{Simulations performed with different numerical grids in Icarus. The blue line corresponds to in-situ data observed at the ACE spacecraft. On the left panel the radial velocity, number density, and logarithm of plasma-$\beta$ are plotted. The right panel shows the magnetic field components and the total magnetic field. The shock arrival time, the magnetic ejecta leading edge (ME LE), and the magnetic ejecta trailing edge (ME TE) are indicated in the catalog from \citet{winslow2015} and are plotted with black, red, and green dashed lines, respectively. Equidistant grid simulations on the low and medium-resolution grids are given in orange and green. The simulations performed on the stretched grid with AMR levels 2, 3, and 4 are given in red, purple, and brown, respectively.
    }
    \label{fig:2013_ace_v_b_timeseries}
\end{figure*}

Figure~\ref{fig:2013_ace_v_b_timeseries} shows the plasma and magnetic field parameters at ACE. The color choices are similar to the previous figure, the observed data is plotted in blue, and the arrival of the disturbances, ME LE and ME TE, are plotted with black, red, and green vertical dashed lines, respectively. 
The left panel focuses on plasma parameters and plasma$-\beta$ in the simulation. when considering the arrival time the speed profiles are modeled well, but the number density peaks slightly earlier than in the observation data. The shock in the speed values is steeper in AMR level 3 and 4 simulations than in the equidistant medium-resolution simulation. The value of the plasma beta indicates that the magnetic cloud arrival is in good agreement with the observed data. In this panel, we can see the two-peak structure in AMR level 4 simulation, which can be connected to the peaks at the arrival of the disturbances at Earth and the arrival of the magnetic cloud. The two peaks are also present in the equidistant medium-resolution simulation, but the first peak is stronger than the second one, in this case. 

On the right panel of Figure~\ref{fig:2013_ace_v_b_timeseries}, we can see that 
the simulated $B_x$ component of the magnetic field closely resembles the profiles in the observed data. In the case of the $B_y$ and $B_z$ components, the overall profiles are similar in the modeled and observed data, but the profiles in the magnetosheath and beginning of the ME seem to be compressed in the modeled data, not catching the exact profile of the observed data. In the simulation study, it was observed that the magnetic cloud arrived a bit later in the equidistant medium-resolution simulation. This delay could be explained by the fact that the shock arrival in the radial velocity panel was less steep, indicating a gradual increase compared to other simulation runs. This indicates that the front of the CME is more diffuse in the equidistant simulations than in the other simulations and the sheath region is modeled more poorly. When considering the total magnetic field panel, the increased strength in the sheath region can be distinguished well in AMR level 3 and 4 simulations, with AMR level 4 results being slightly more similar to the observed data. The stronger magnetic field region arrives later in the equidistant medium-resolution simulation and appears to be narrower than in the AMR simulations and the observed data. The strength of the total magnetic field is underestimated in the simulation results.

\begin{figure*}
    \centering
    \includegraphics[width=0.33\textwidth]{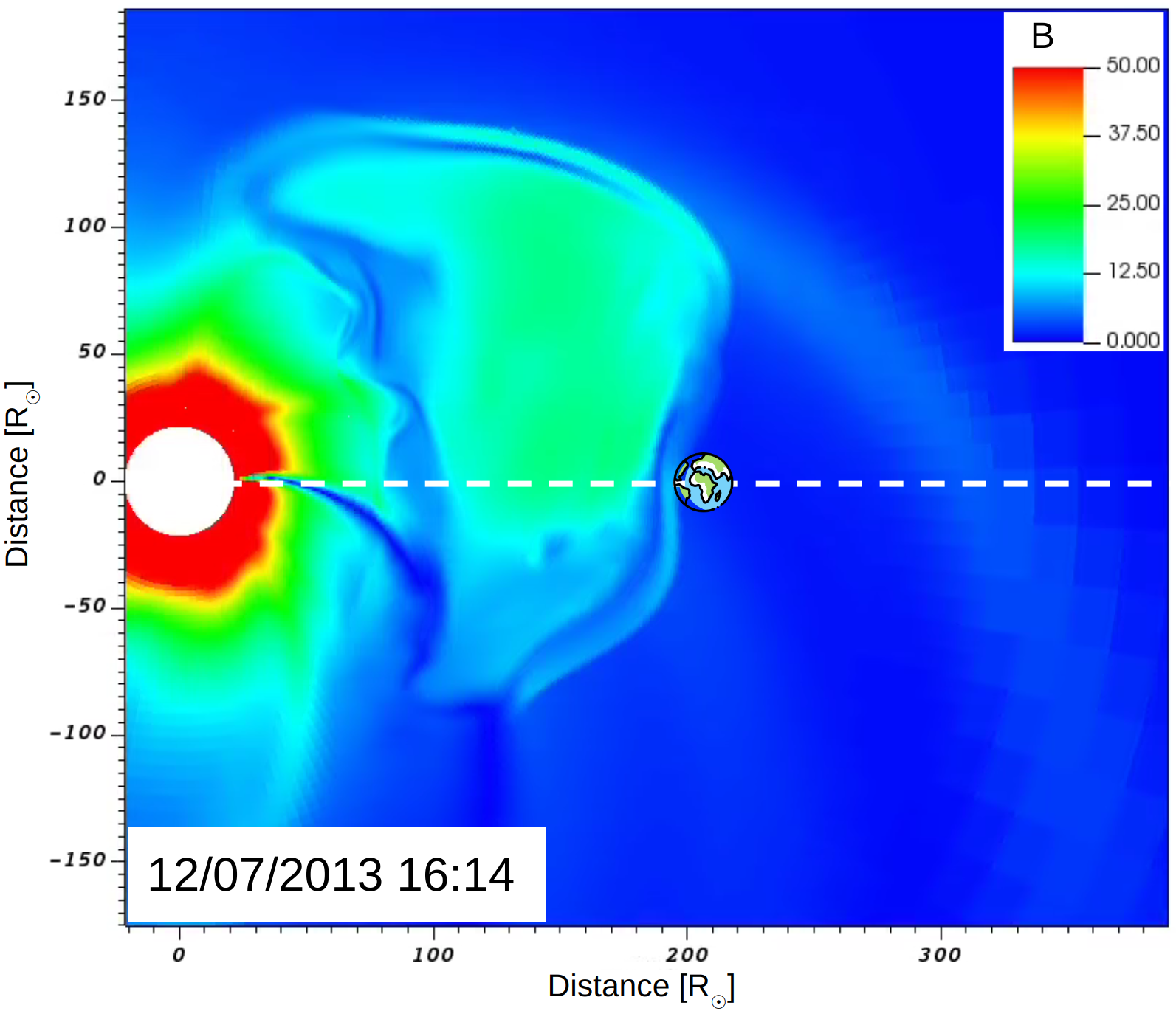}
    \includegraphics[width=0.33\textwidth]{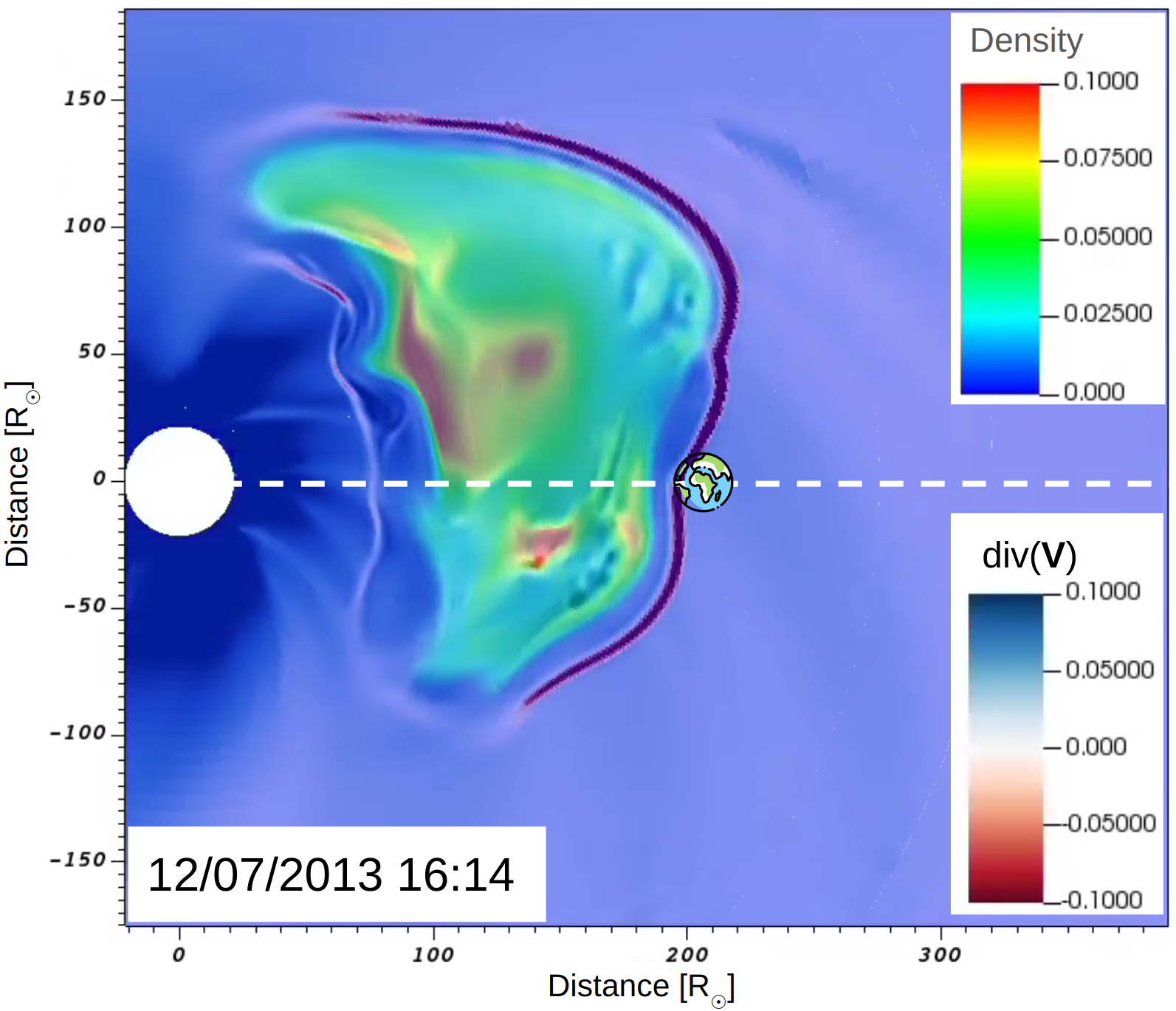}
    \includegraphics[width=0.33\textwidth]{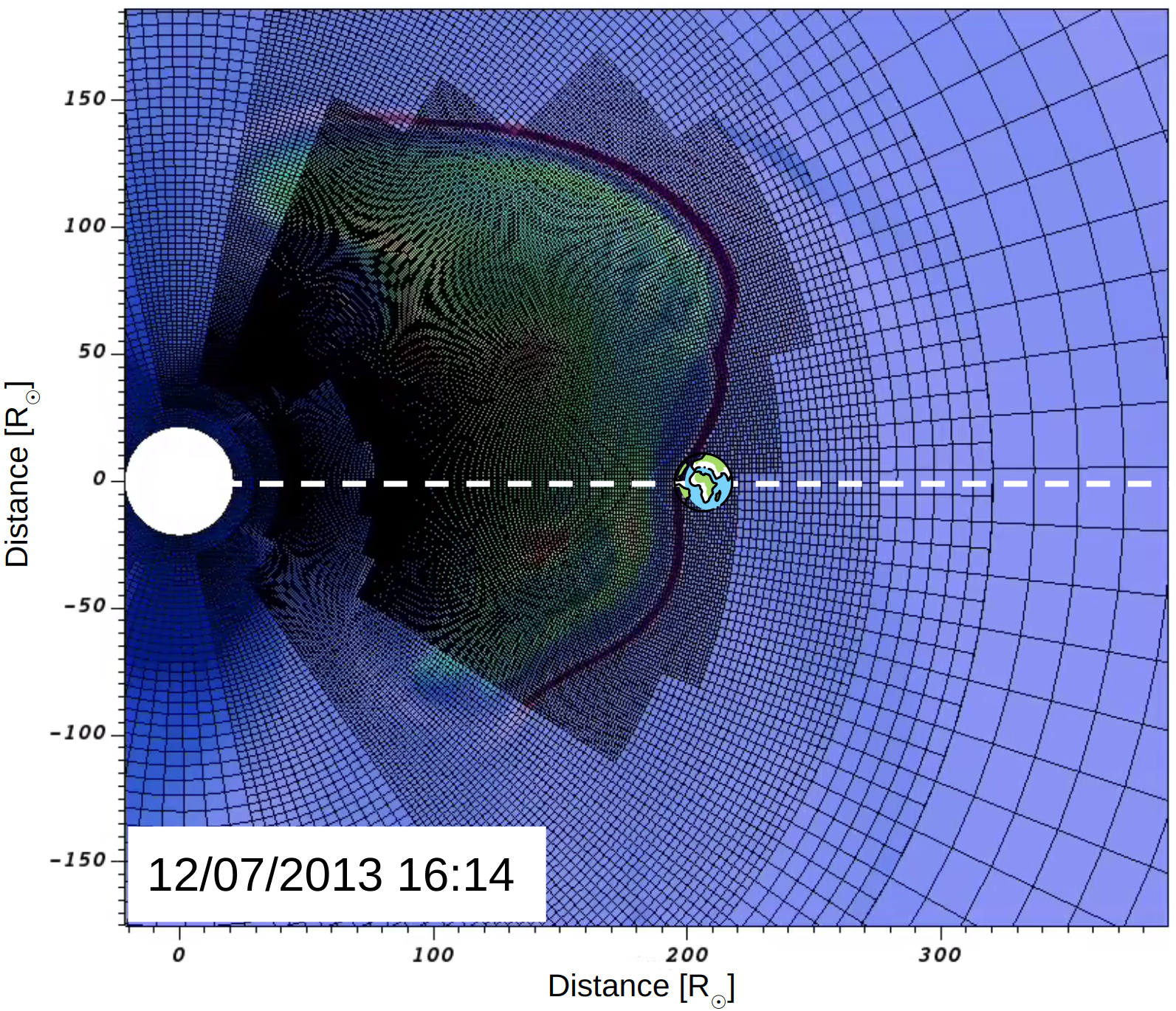}
    \caption{The equatorial plane in Icarus at the arrival time of the disturbance at 1~\AU. The distance from the Sun is indicated on the horizontal and vertical axes. The left panel shows the slice coloured with total magnetic field values. The medium figure represents the plasma density inside the CME overlaid with the divergence of velocity values. The right panel shows the medium panel with the computational grid plotted on top. The frame is rotated so that Earth is fixed at 0 longitude. The white dashed line is along the Sun-Earth line.}
    \label{fig:2013_icarus_snapshots}
\end{figure*}

\begin{figure*}
    \centering
    \includegraphics[width=0.33\textwidth]{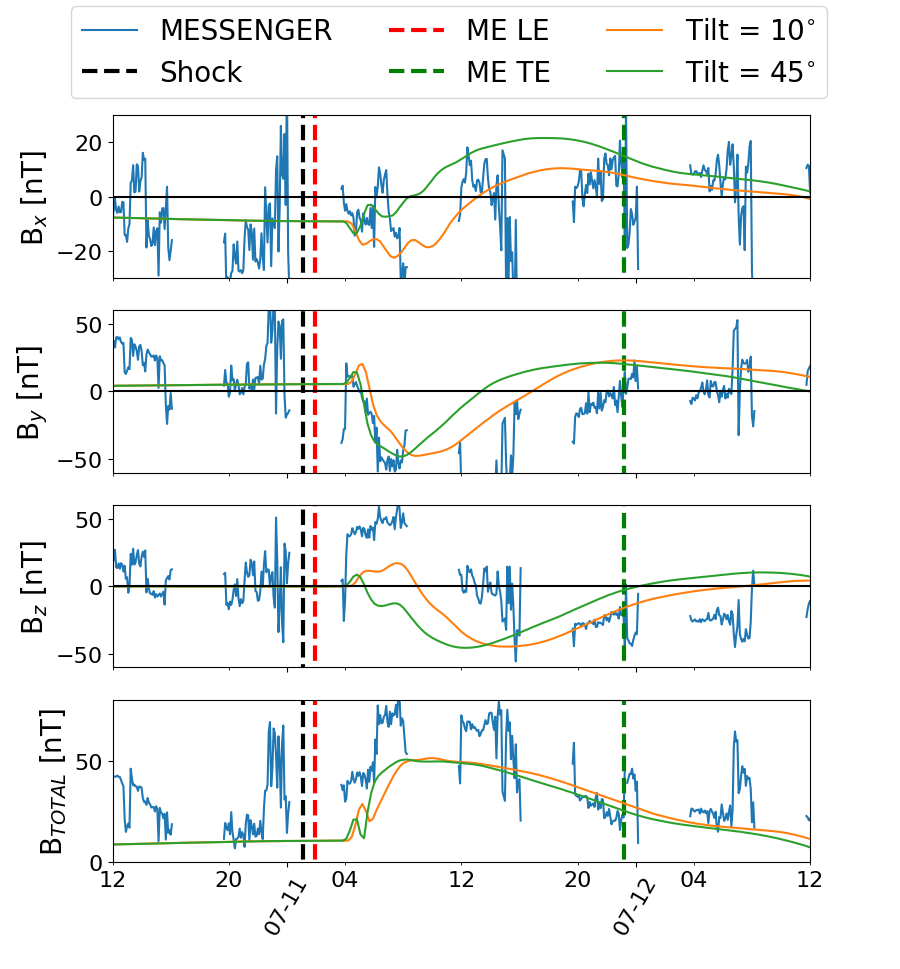}
    \includegraphics[width=0.33\textwidth]{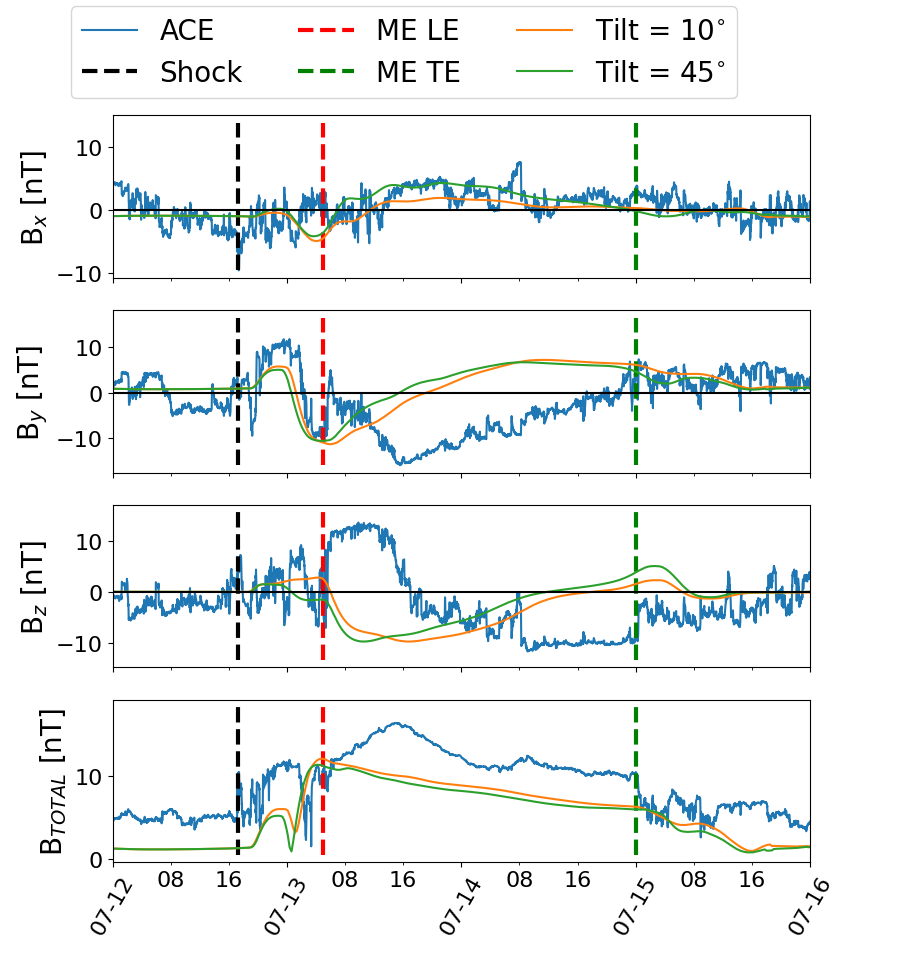}
    \includegraphics[width=0.33\textwidth]{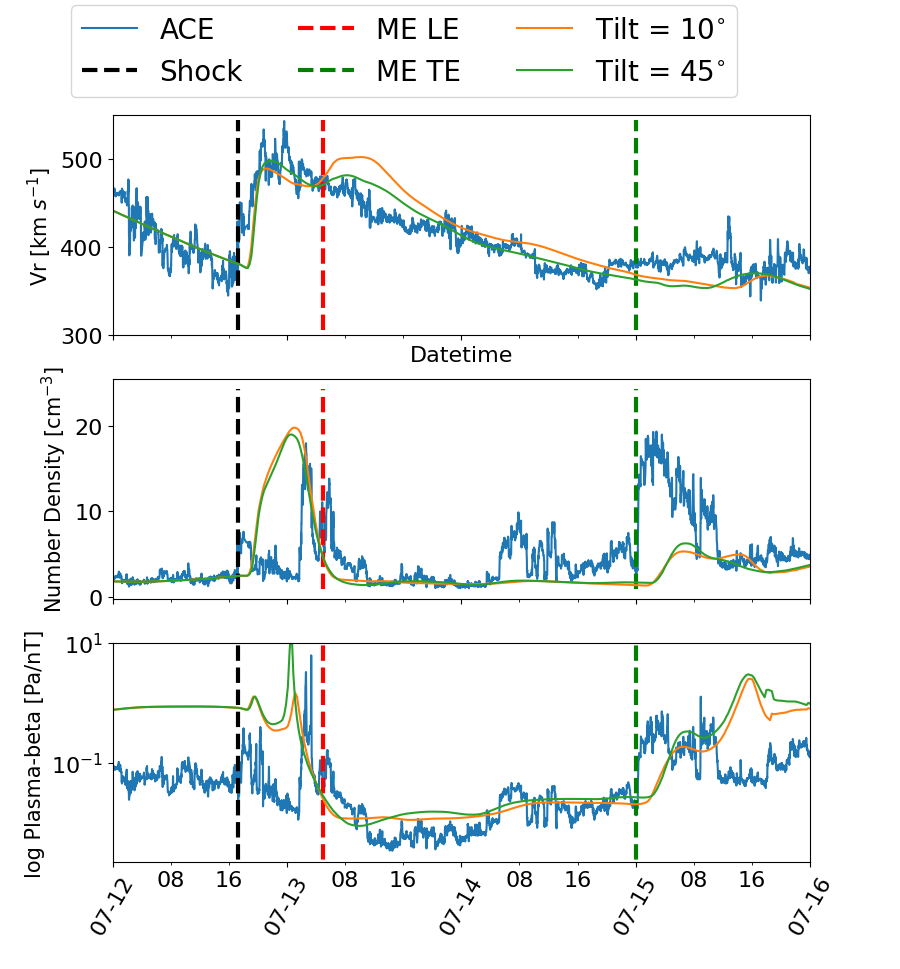}
    \caption{Observations at Mercury and ACE are plotted in blue similar to Figures~\ref{fig:2013_messenger_b_timeseries} and \ref{fig:2013_ace_v_b_timeseries}. The left panel represents magnetic field components at MESSENGER and the middle and right panels at ACE. The orange curve corresponds to the same parameter set-up as reported previously, with the tilt$\;= 10^\circ$ and green corresponds to the same parameter set, apart from tilt$\;= 45^\circ$. The simulations are performed on an AMR level 4 grid. }
    \label{fig:2013_tilt_comparisons}
\end{figure*}

\begin{table}[t!]
\caption{The wall-clock times for the performed simulations in Icarus. }   
\label{table:run_times_2013}   
\centering            
\begin{tabular}{c c c c c}         
\hline
Low$_\text{EQ}$& Middle$_\text{EQ}$ & AMR 2 & AMR 3 & AMR 4  \\ 
\hline\hline
  7m 55s & 1h 13m 19s & 4m 20s & 7m 49s& 34m 32s  \\ \hline                                 
\end{tabular}
\tablefoot{
 The runs are performed with different computational grids on 4 nodes with 2 Xeon Gold 6240 CPUs@2.6 GHz (Cascadelake), 18 cores each, on the Genius cluster at KU Leuven. Low$_\text{EQ}$ and Middle$_\text{EQ}$ correspond to simulations on the equidistant low and medium-resolution grids.  
}
\end{table}

Overall, the CME interaction at Mercury and Earth is modeled well, considering that the arrival time and the strengths of the variables and profiles are in good agreement with the observed data at Mercury by the MESSENGER spacecraft and near Earth by the ACE spacecraft. \cite{Grison2018} reported that no significant acceleration or deceleration was calculated for this CME event. From the simulations, we could also see that at both locations it arrived close to the arrival time reported in \cite{winslow2015} and \cite{Salman2020}. When comparing the magnetic field components at two different locations, we can see no significant change in the orientation of the magnetic field configurations, which indicates that the two spacecraft passed through the CME at similar locations and no significant rotation and deflection took place from the orbit or Mercury to the orbit of Earth. 

When considering different computational grids for modeling the solar wind and the CME, we can see that near Mercury, the equidistant medium-resolution simulation result is more similar to the AMR level 3 simulation results, when comparing the resolved small variations in the time-series. The AMR level 4 simulation resolves the structure of the magnetic field upon its arrival in more detail than the other simulations and performs better than the equidistant medium-resolution simulation. 

Figure~\ref{fig:2013_icarus_snapshots} shows the results from the Icarus simulation performed on the stretched grid in combination with AMR refinement level 4. The snapshots are taken in the equatorial plane at the arrival time of the CME shock at 1~\AU. The white dashed line is along the Sun-Earth line and the Earth icon is placed at the location of Earth in the given snapshot. The frame is rotated so that Earth is fixed at $0^\circ$ longitude. The left panel represents the magnetic field strength with the given color bar in [nT]. The middle panel represents the CME density values, saturated to $(0, 0.1)$ in the code units, for better visibility.
The density profile is overlaid by the divergence of $\Vec{V}$ values, showing the compression regions in the domain. The values are saturated to $(-0.1, 0.1)$ to better distinguish the compression regions, given in red. Thus, from this panel, we can see the whole structure, the red shock front, and the CME interior behind it. The right panel represents the same values plotted as in the middle panel but with the computational grid plotted on top. The same portion is plotted from the equatorial plane without zooming for consistency. The dense region corresponds to AMR level 4 covering most of the CME interior and the CME shock front, as indicated in the AMR refinement criterion. 

Finally, we performed a simulation with the tilt that was calculated from the PIL line at the solar photosphere and compared it to a simulation using the tilt deduced from the in-situ measurement by \cite{Palmerio2018}. Figure~\ref{fig:2013_tilt_comparisons} shows the time series at Mercury (left panel) and at ACE (middle and right panels). Both simulations were performed on the stretched grid with 4 AMR levels, as these simulations resolve more small-scale structures in time series than others. The orange curve corresponds to the parameter setup given in Table~\ref{table:2013_event_parameters}, with tilt$\;= 10^\circ$, while the green curve corresponds to the simulation with tilt$\;= 45^\circ$. At Mercury, we can see, the $B_x$ component is overestimated in the simulation with tilt$\;=45^\circ$. The $B_y$ component is comparable to the standard tilt simulation, however, the magnetic cloud seems less extended.

The $B_z$ component is underestimated compared to the simulation results with tilt$\;=10^\circ$. Overall, at Mercury, the simulation with tilt$\;=10^\circ$ produces results that are more similar to the observed data than with tilt$\;=45^\circ$. At ACE, the profiles corresponding to the simulations with tilt fixed to $10^\circ$ and $45^\circ$ are more similar and no significant difference can be spotted. When comparing the magnetic field component profiles at Mercury and ACE, we can see that there is a larger difference in the case of the simulation with tilt$\;=45^\circ$, especially in the $B_y$ and $B_z$ components, which can be explained to the rotating feature of the spheromak model in the simulation \citep{Asvestari2022, Maharana2023,Sarkar2024}. This indicates that the rotation of the flux-rope occurred in the solar corona or the lower heliosphere, below $\sim 0.35$~\AU, since the magnetic field components are better recovered by the simulation with the tilt fixed to 10$^\circ$ at both locations in the heliosphere.

\begin{figure*}
    \centering
    \includegraphics[width=0.33\textwidth]{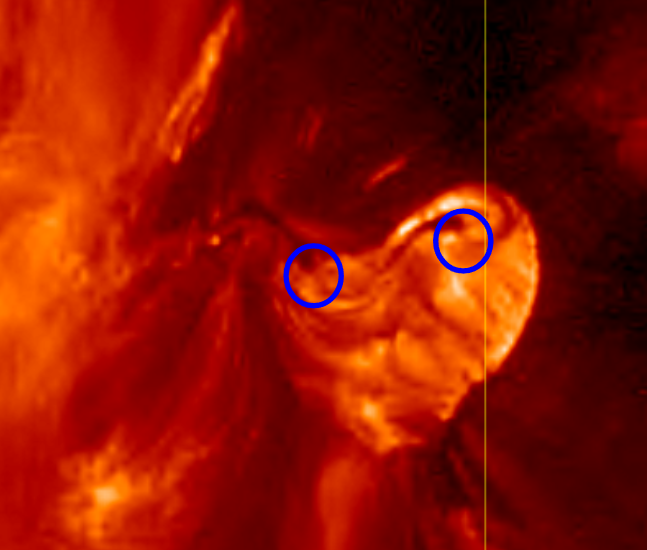}
    \includegraphics[width=0.33\textwidth]{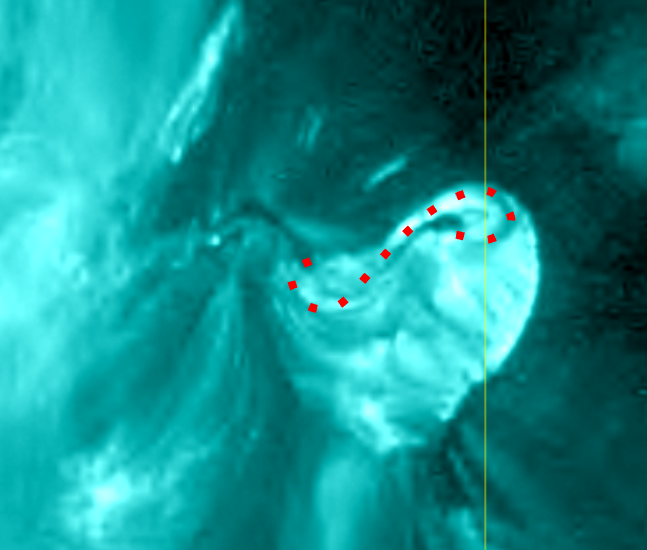}
    \includegraphics[width=0.33\textwidth]{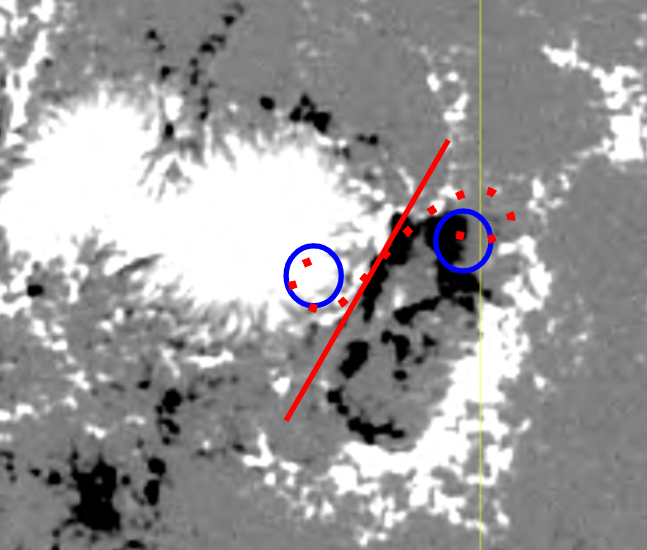}
    \caption{The images taken from JHelioviewer correspond to the source of the CME related to the eruption on SOL2014-02-16T09:22:11. The left panel shows the image of the solar atmosphere in Solar Dynamics Observatory/Atmospheric Imaging Assembly 304 \textup{~\AA} wavelength, where the circles identify the footpoints. The middle panel shows the solar corona in SDO/AIA 131 \textup{~\AA} wavelength, with a dotted sigmoid shape plotted on top. The right panel shows the HMI magnetogram with the red line corresponding to the polarity inversion line (PIL) together with the dotted sigmoid and the footprints plotted on the left and middle panels.}
    \label{fig:2014_source}
\end{figure*}

Table~\ref{table:run_times_2013} shows the wall-clock times for the different Icarus simulations all performed on 4 nodes on the Genius cluster at KU Leuven. Each node contains 2 Xeon Gold 6240 CPUs@2.6 GHz (Cascadelake), 18 cores each. The equidistant medium-resolution simulation took the longest as expected, $\sim 1\;$h 14 min. The AMR level 4 simulation, which produced comparable or better results when looking at time series data, took only $\sim 35$ minutes, which is $\sim 2.1 $ times faster than the equidistant medium-resolution simulation. The AMR level 3 simulation took under 8 minutes, similar to low-resolution simulation while showing considerably more small-scale features in the time-series data. Thus, performing simulations with the combined refinement criterion, which aims at a large area is still more advantageous than using the standard equidistant grid without optimization. 

\section{CME event II: February 16, 2014} \label{2014_february}

\subsection{Observations of the studied CME}

For the second analysed event we chose the CME which occurred on February 16, 2014, at 10:24 UT, as reported by \cite{winslow2015} and \cite{Salman2020}. The CME was observed at MESSENGER on February 17, 2014, at 04:17, and at ACE on February 19, 2014, at 03:48. 
The corresponding spacecraft longitudinal separation was $4.8^\circ$. The source was identified in the solar atmosphere, corresponding to the active NOAA 11977, and the M1.1 class flare occurred at 09:00 UT with the peak at 09:26 am with coordinates S10E00. The snapshots of the region before the eruption are given in Figure~\ref{fig:2014_source}. The left figure shows the AIA 304\textup{~\AA} wavelength filter, where possible foot-points are encircled in blue. The middle panel shows AIA 131\textup{~\AA} to distinguish the sigmoid structure better. The approximated S-shaped sigmoid is plotted with the red dotted line at the eruption site. The sigmoid's orientation means that the eruption's chirality is positive. Therefore, the flux-rope magnetic field configuration is right-handed. The right panel shows the HMI magnetogram with the PIL plotted on top, the approximated sigmoid structure, and the footprint sites for the whole picture. The PIL suggests that the axial direction is northwest, and the flux rope is WNE/SWN type. The tilt calculated from the PIL is $\sim 59.6^\circ$.

\begin{figure}[ht!]
    \centering
    \includegraphics[width=0.5\textwidth]{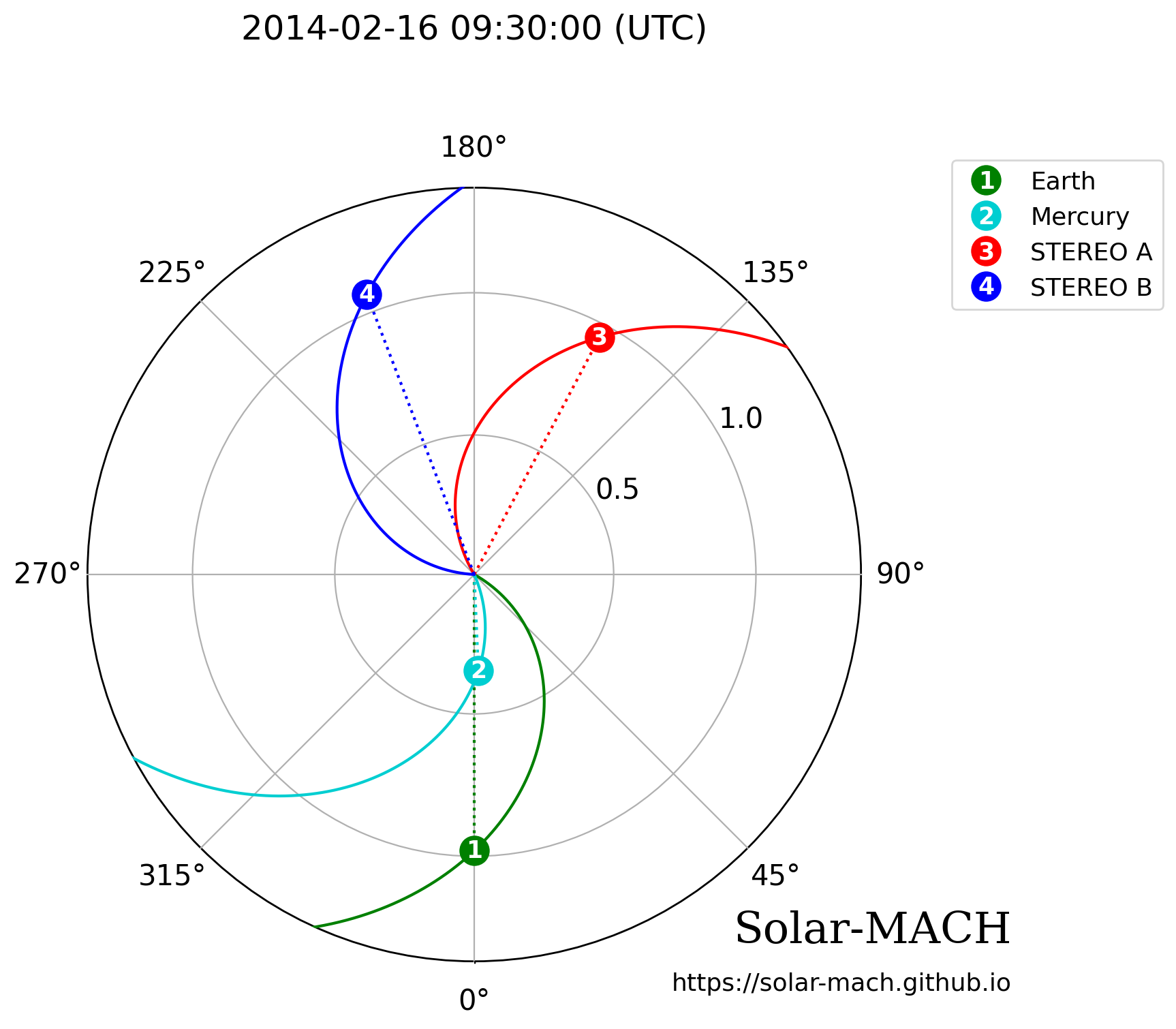}
    \caption{The alignment of the Mercury and Earth at the time of the CME eruption. The locations of STEREO-A and STEREO-B are also plotted. The figure is generated with Solar-MACH \citep{solarmach}.}
    \label{fig:2014_mercuryearth_aligneent}
\end{figure}

\begin{figure}[ht!]
    \centering
    \includegraphics[width=0.4\textwidth]{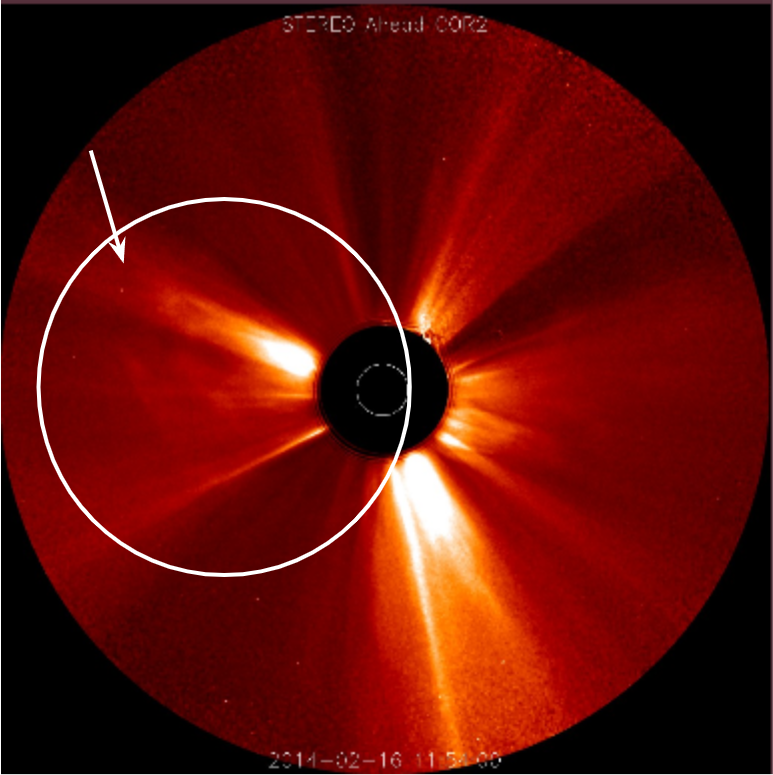}
    \includegraphics[width=0.4\textwidth]{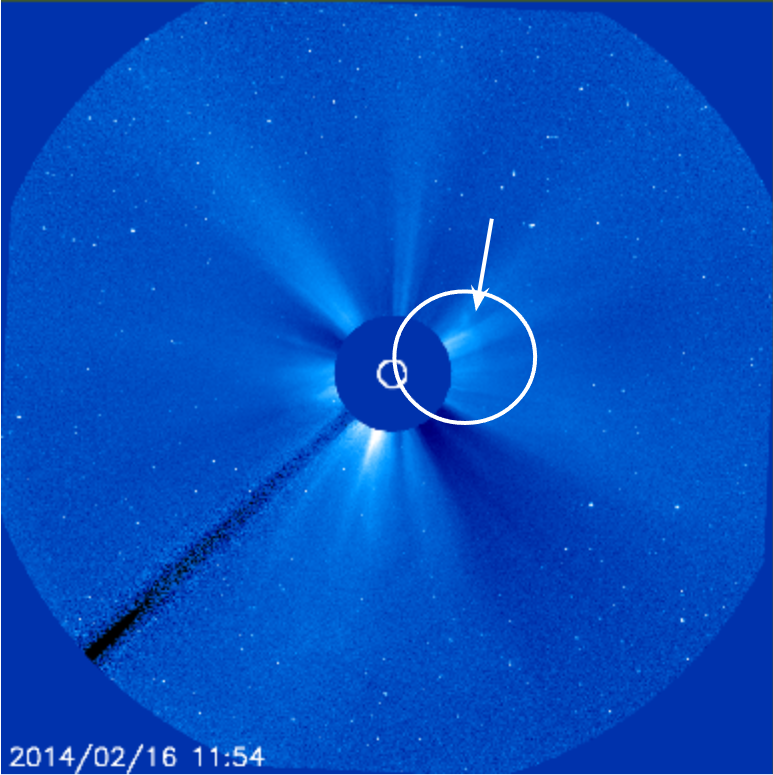}
    \caption{The white-light images seen from STEREO-A Cor2 and SOHO/LASCO Cor3 $\sim 2.5$ hours after CME$_C$ eruption. The figure is generated with the Stereo CME Analysis Tool.}
    \label{fig:2014_sta_lasco}
\end{figure}

\begin{table}[t!]
\caption{The CMEs registered at Earth in the catalog by \citet{richardson_catalog}.
}  

\label{table:3cmes}   
\centering            
\begin{tabular}{c c c c}         
\hline
CME event &Disturbance start & ME LE & ME TE  \\ 
\hline\hline
 CME$_P$ & 02/18 06:40 & 02/18 15:00 &02/19 07:00    \\ 
 CME$_C$  & 02/19 03:48 & 02/19 12:00  &02/20 03:00  \\      
 CME$_A$  & 02/20 03:18 & 02/21 02:00  &02/22 12:00    \\ \hline    
\end{tabular}
\tablefoot{
 The CME$_C$ corresponds to the CME event chosen from the \citet{winslow2015} catalog. The CME$_P$ and CME$_A$ indicate the disturbances arriving previously and after CME$_C$. The events are from the year 2014.
 
}

\end{table}

\begin{figure*}[ht!]
    \centering
    \includegraphics[width=0.32\textwidth]{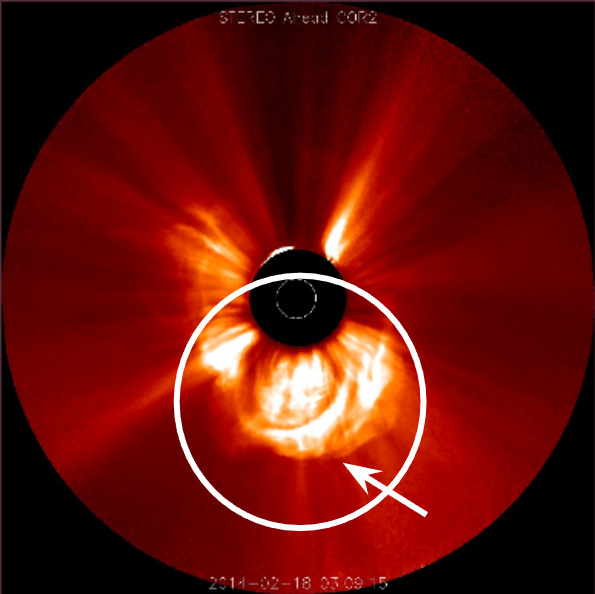}
    \includegraphics[width=0.32\textwidth]{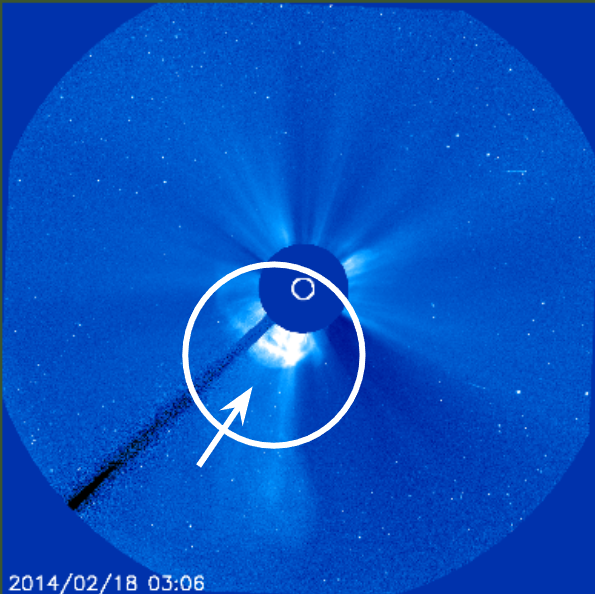}
    \includegraphics[width=0.32\textwidth]{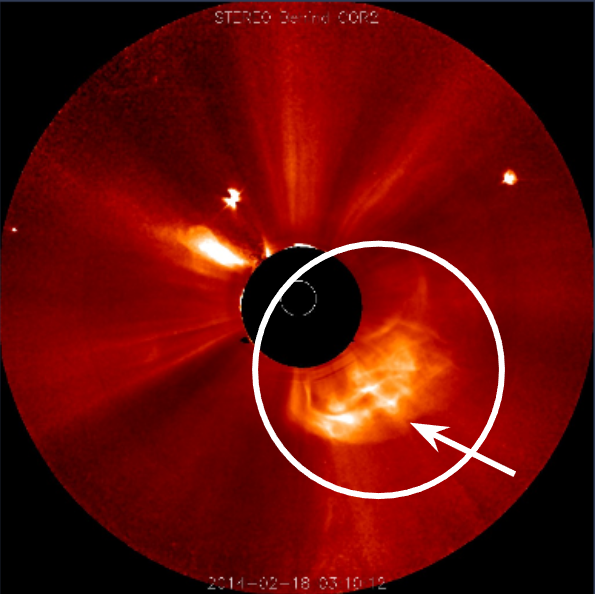}
    \caption{The white-light images seen from STEREO-A Cor2, SOHO/LASCO Cor3 and STEREO-B C2 $\sim 1.5$ hours after the CME$_A$ eruption. The figure is generated with the Stereo CME Analysis Tool.}
    \label{fig:2014_3rd_sta_lasco_stb}
\end{figure*}

\begin{figure*}
    \centering
    \includegraphics[width=0.33\textwidth]{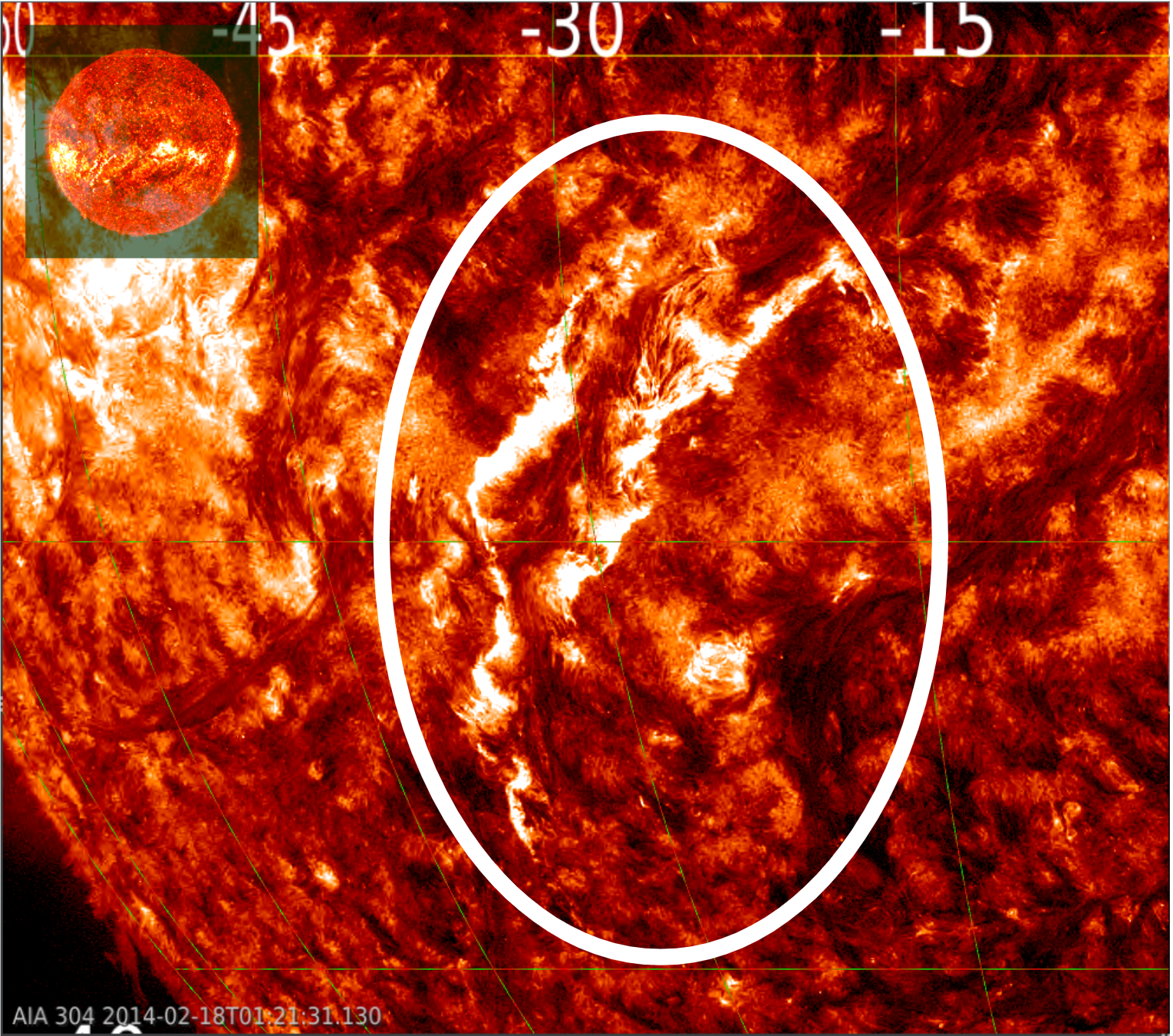}
    \includegraphics[width=0.33\textwidth]{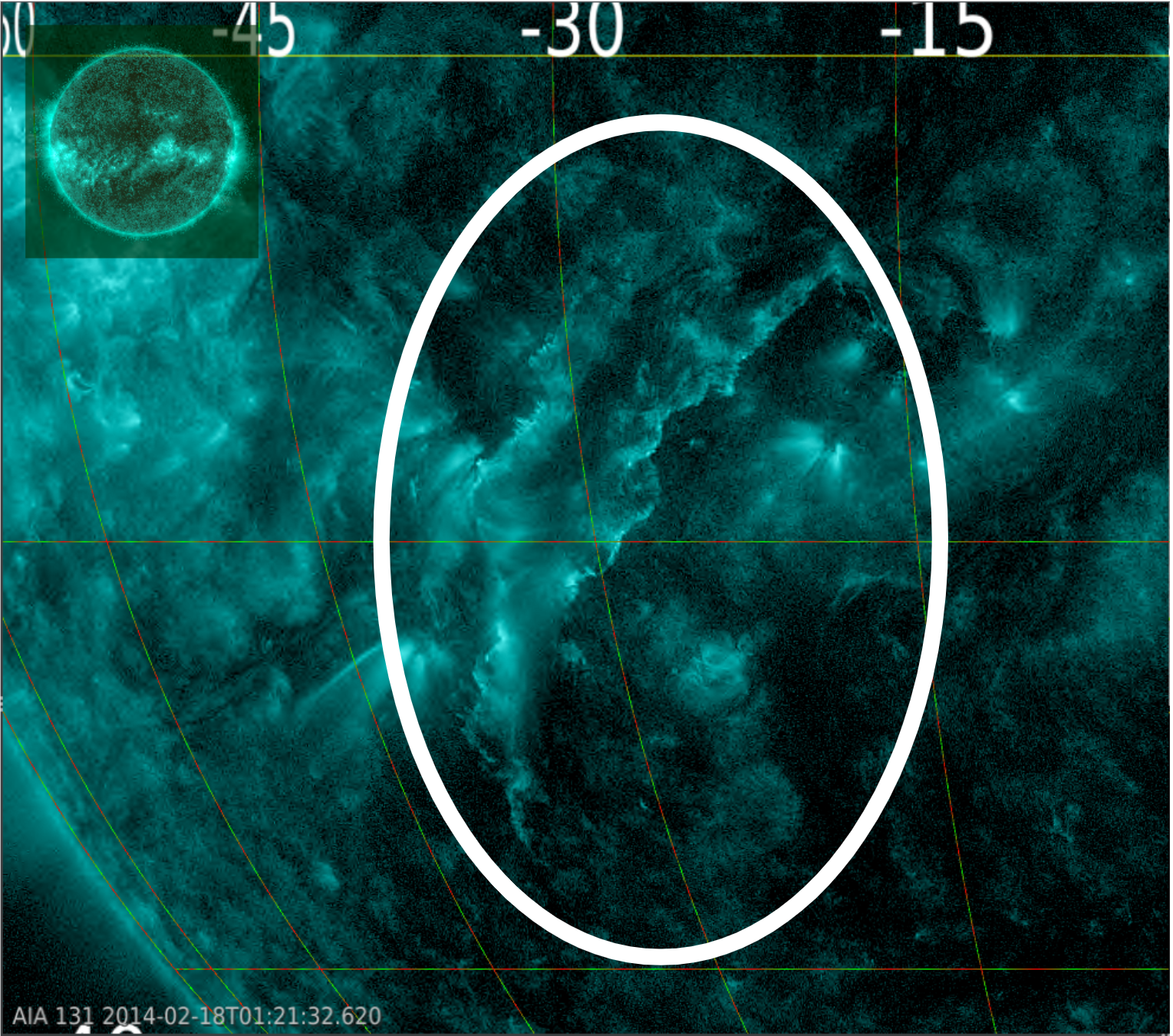}
    \includegraphics[width=0.33\textwidth]{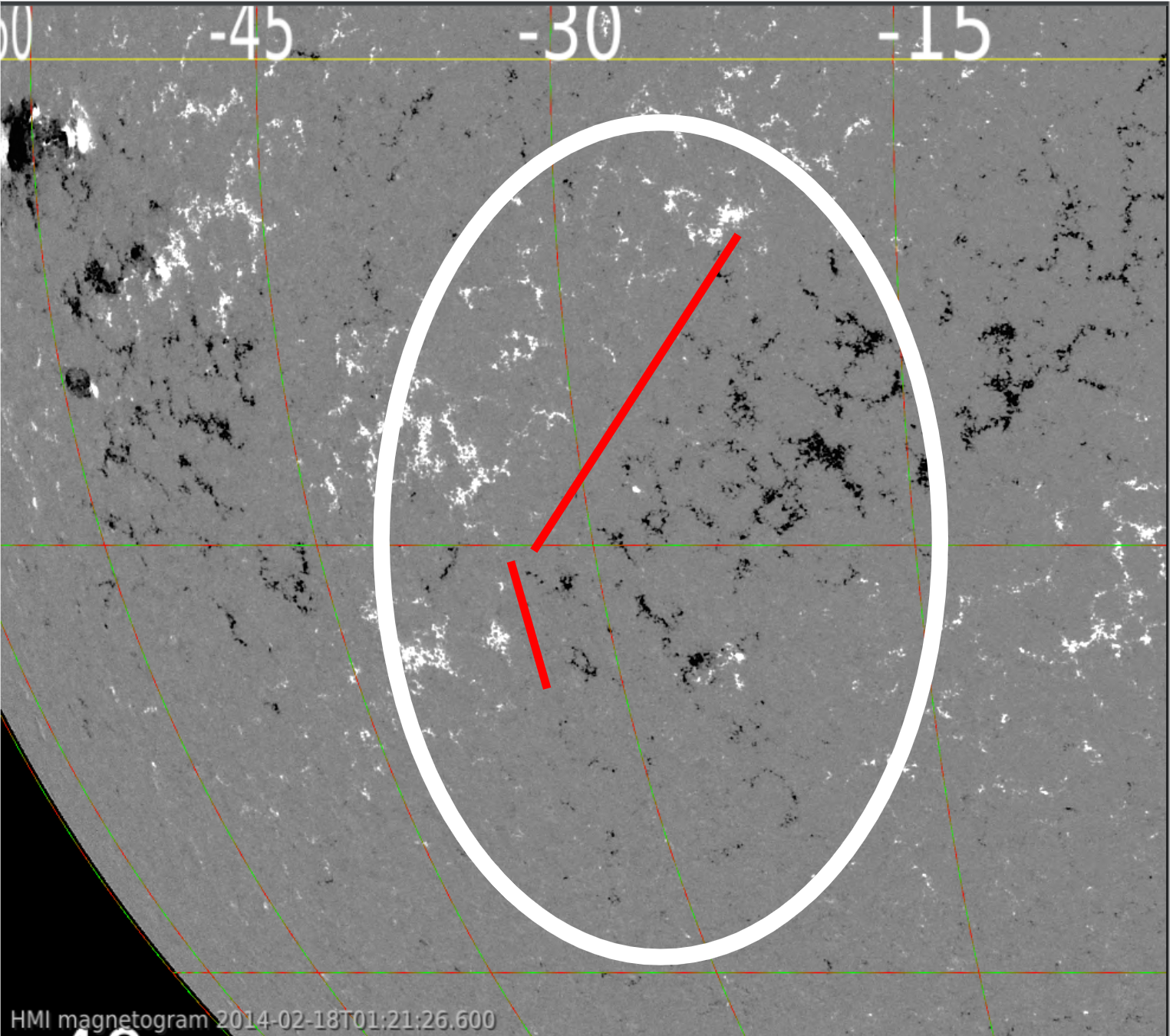}
    \caption{The images taken from JHelioviewer correspond to the source of CME$_A$ event before the eruption on 2014-02-18T01:21:26. The left and right panels show the images of the solar atmosphere in Solar Dynamics Observatory/Atmospheric Imaging Assembly 304\textup{~\AA} and 131\textup{~\AA} wavelengths, respectively. The white circle highlights the filament area. The right panel shows the HMI magnetogram with the red lines corresponding to the polarity inversion lines (PIL) found in the filament region. }
    \label{fig:2014_3rd_CME_source}
\end{figure*}

The CME event was observed by MESSENGER and ACE, and their locations, together with Stereo-A and Stereo-B, are plotted in Figure~\ref{fig:2014_mercuryearth_aligneent}. The angular separation between Mercury and Earth is slightly larger (by $1.7^\circ$) here than for the previous CME event studied. The CMEs were not well distinguishable in the available coronagraph images. Nevertheless, the signatures can be seen in Stereo-A C2 and SOHO/LASCO C3 images, given in Figure~\ref{fig:2014_sta_lasco}. The CME is encircled in white, and the leading parts are indicated by a white arrow. The white-light images from Stereo-B did not contain good enough images to distinguish CME contribution. Therefore, they are not plotted. The snapshots in Figure~\ref{fig:2014_sta_lasco} are taken after $\sim2.5$ hours since the launch of the CME from the solar atmosphere. Therefore, the CME has already expanded in the C2 image and can already be seen in C3. Since the CME can not be distinguished very well, obtaining parameters from StereoCAT and GCS fitting was challenging. The obtained parameters were used as initial guesses and further optimized when comparing the modeled data to the observed data. 

\subsection{Three interacting CMEs}

However, when injecting only this single CME in the heliosphere, the synthetic time series poorly matched the in-situ data, missing prominent features in the data. The catalog by \citet{richardson_catalog} indicates CMEs registered at Earth (L1) before and after this CME. According to this catalog, the registered disturbances on Earth are given in Table~\ref{table:3cmes}. CME$_P$, CME$_C$, and CME$_A$ stand for the previous CME, the current CME, and the CME that arrives after, respectively. From this table, we can see that the arrival of the disturbance for the second CME is before the ME of the first one has passed Earth. Therefore, there is a CME-CME interaction occurred before 1~au. 

The third CME arrival time is right after the second one (CME$_C$) 
has passed Earth. This indicates a second CME-CME interaction along the way, and the third CME compressed the second CME. Since with the isolated second CME modeling the time series did not replicate any of the prominent features in the observed data, we also modeled the other two CMEs, resulting in a heliospheric simulation with three consecutive CMEs being injected from the inner boundary.

The source of the preceding CME was not identified, which was also reported in the study by \citet{Wang2018}. We estimated the CME parameters to model it at 0.1~au from the in-situ data to reconstruct the plasma conditions before the arrival of CME$_C$. The CME must have had low speed, as upon its arrival at Earth, the speed increases suddenly from $\sim350\;$km s$^{-1}$ to $\sim420\;$km s$^{-1}$ (Figure~\ref{fig:2014_ace_v_timeseries}). The CME speed profile is flat. The ME arrives 8.5 hours later, and it takes 15 hours to pass ACE according to Table~\ref{table:3cmes}. However, the in-situ data reveal that after the arrival of the magnetosheath of the second CME, there is a slight variation in the plasma$-\beta$ profile. But the ME ended passing ACE at 15:00 am from Figure~\ref{fig:2014_ace_v_timeseries}, meaning the passage took 21 hours. The magnetic field strength does not significantly increase since it remains under $10\;$nT throughout the ME passage while it reaches $\sim20\;$nT in the preceding sheath. Considering the orientation of the magnetic field components, a left-handed CME configuration was assumed with $0^\circ$ tilt. 

\begin{figure*}
    \centering
    \includegraphics[width=0.48\textwidth]{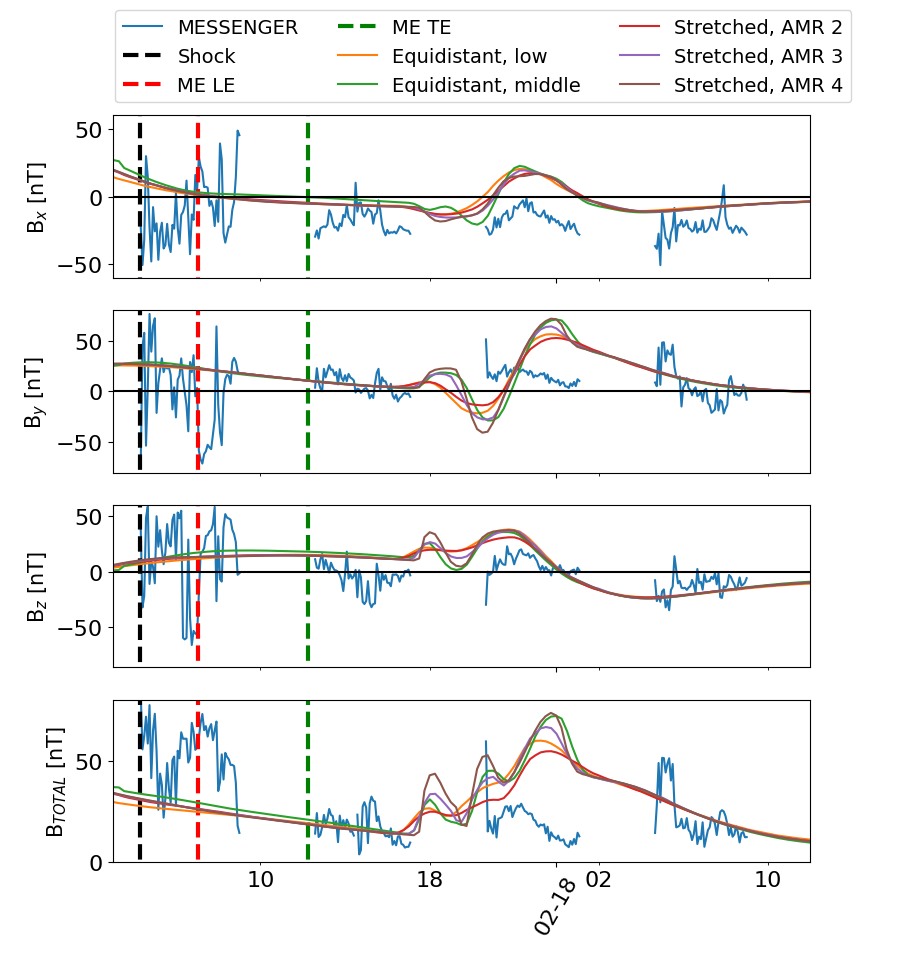}
    \includegraphics[width=0.48\textwidth]{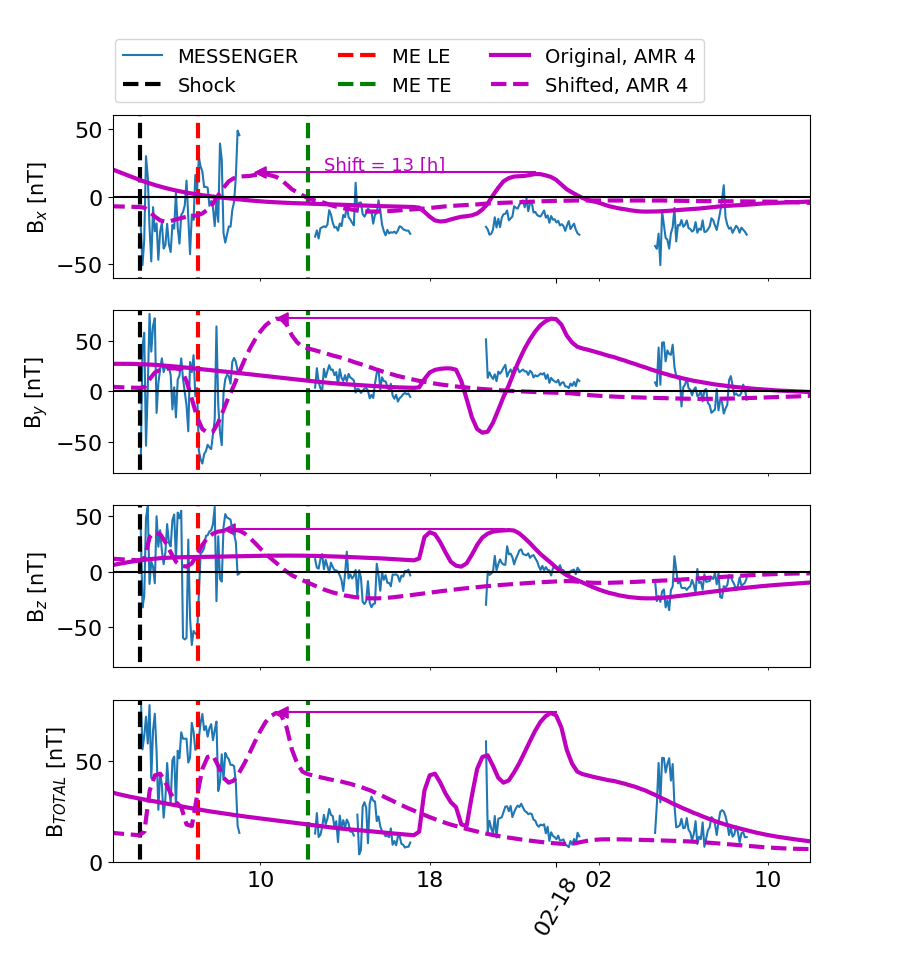}
    \caption{Simulations performed with different numerical grids in Icarus. The blue line corresponds to in-situ data observed at the MESSENGER spacecraft. The black, red, and green vertical dashed lines indicate the shock arrival time, the magnetic ejecta leading edge (ME LE), and the magnetic ejecta trailing edge (ME TE), respectively. The left figure shows magnetic field components. Equidistant grid simulation results with low and medium-resolution grids are given in orange and green. The results of the simulations performed on the stretched grid with the combination of AMR levels 2, 3, and 4 are given in red, purple, and brown, respectively. The right figure shows the shift of the modeled data in time with 13 hours to project the profiles during the CME interaction at MESSENGER. }
    \label{fig:2014_messenger_b_timeseries}
\end{figure*}
 
 On the other hand, the third CME, CME$_A$, was found in The Space Weather Database Of Notifications, Knowledge, Information (DONKI) catalog of CCMC with estimated direction, size, and propagation speed. According to the parameters reported in the DONKI catalog, the CME was launched at $(\phi, \theta) = (-29^\circ, -19^\circ)$ in HEEQ coordinates with a speed of $V_r = 600\;$km s$^{-1}$ and a half angular width $\omega /2 = 53^\circ$. The CME reached 0.1~au, the inner heliospheric boundary, on 2014-02-18 at 07:32 UT. Since these parameters were obtained for the simple hydrodynamics plasma CME model, the parameters were fitted with the StereoCAT online tool and GCS 3D reconstruction and further optimized in the simulations. The aim of including the third CME in the simulation is to investigate its effect on the main CME we are modeling. The CME was visible at STA, SOHO/LASCO, and STB, as shown in Figure~\ref{fig:2014_3rd_sta_lasco_stb}. As can be seen from these figures, CME$_A$ is large and has a high speed since the snapshots are taken only $\sim1.5\;$hours after the CME launch. After performing the fitting of the CME with the StereoCAT online tool, the obtained $(\phi, \theta) = (-10^\circ, -36^\circ)$ in HEEQ coordinates with the speed of $V_r = 950\;$km s$^{-1}$ and half angular width $\omega /2 = 43^\circ$. We located the source on the solar photosphere to identify the orientation of the flux rope and estimate the tilt for the input spheromak CME model. Figure~\ref{fig:2014_3rd_CME_source} shows the source region before the eruption. In all three figures, the location of the eruption is indicated with a white circle. The left figure is made through the AIA 304\textup{~\AA} wavelength filter. The middle figure shows the solar corona through the AIA 131\textup{~\AA} filter. The right figure shows the HMI magnetogram. The possible PIL locations are over-plotted with red solid lines at the possible eruption site. The exact CME source was hard to identify since the eruption can only be seen near the limb. The images in the AIA 131\textup{~\AA} and AIA 304\textup{~\AA} filters are blurry, and it is hard to distinguish the pre-eruptive structures to identify the tilt and exact helicity of the flux-rope. We started with the PIL tilt shown in the figure and further optimized it in the simulations. Since the goal of including the third CME was only to check its effect on the second one and whether it compresses the previous one, further identification of the source of the third CME was not pursued. The obtained parameters are given in Table~\ref{table:2014_event_parameters}, which were used for modeling CMEs in the heliosphere.

\begin{table}[t!]   
\centering  
\caption{The parameters for the input for the spheromak CME model for the three CMEs for February 16, 2014.}   
\begin{tabular}{c c c c }         
\hline 
Variable & CME$_P$ & CME$_C$ & CME$_A$ \\ 
\hline  \hline        
    t$_{CME}$ & 02-14T12:34 & 02-16T17:34 & 02-18T07:32  \\
    $\theta_{CME}$ & 0$^\circ$ & -5$^\circ$ & -10$^\circ$  \\
    $\phi_{CME}$ &-15$^\circ$ & -25$^\circ$ & -49$^\circ$  \\ 
    r$_{CME}$ [R$_\odot$] & 15.0 & 17.0  & 17.0\\
    v$_{CME}$ [km s$^{-1}$] & 150 & 400 & 600  \\ 
    $\rho_{CME}$ [kg m$^{-3}$] & 3$\times10^{18}$ & 2$\times10^{18}$ & 6$\times10^{17}$ \\ 
    T$_{CME}$ [K] & 8 $\times 10^5$  & 2.0 $\times 10^6$  & 3.8 $\times 10^6$\\
    $\tau_{CME}$ &  0$^\circ$ & -90$^\circ$ & -45$^\circ$ \\ 
    H$_{CME}$ & -1 & 1 & 1 \\ 
    F$_{CME}$ [wB] & 4$\times 10^{13}$ & $ 3.5 \times 10^{13}$ & 6.4$\times 10^{13}$ \\
\hline  
\end{tabular}
\label{table:2014_event_parameters}
\end{table}

\subsection{Comparison of simulations to data}

Figure~\ref{fig:2014_messenger_b_timeseries} shows the time series obtained at MESSENGER. The MESSENGER data are plotted in blue. The data from passages of the magnetosphere are removed as the CME contribution can not be distinguished there from the strong magnetosphere magnetic field. The black, red, and green vertical dashed lines indicate the arrival of the disturbance, ME LE and ME TE at MESSENGER reported in \cite{winslow2015}, respectively. Unfortunately, MESSENGER enters the magnetosphere of Mercury within 2 hours from the arrival of the ME, so more than half of the CME contribution can not be distinguished from the observed data. This increases the ambiguity when comparing the modeled data to the observed data. In the left figure, the equidistant grid simulation results with the low and middle-resolution grids are plotted with orange and green, respectively. The results for AMR level 2, 3, and 4 simulations on stretched grids are plotted with red, purple, and brown colors, respectively. The right-hand side figure plots the AMR level 4 simulation result in the solid magenta line. The synthetic data was shifted artificially by 13 hours back in time to match the arrival time in the observed data to allow better comparison of the magnetic field components to the MESSENGER observations. The shifted synthetic data is plotted with the dashed magenta line. 

\begin{figure*}
    \centering
    \includegraphics[width=0.48\textwidth]{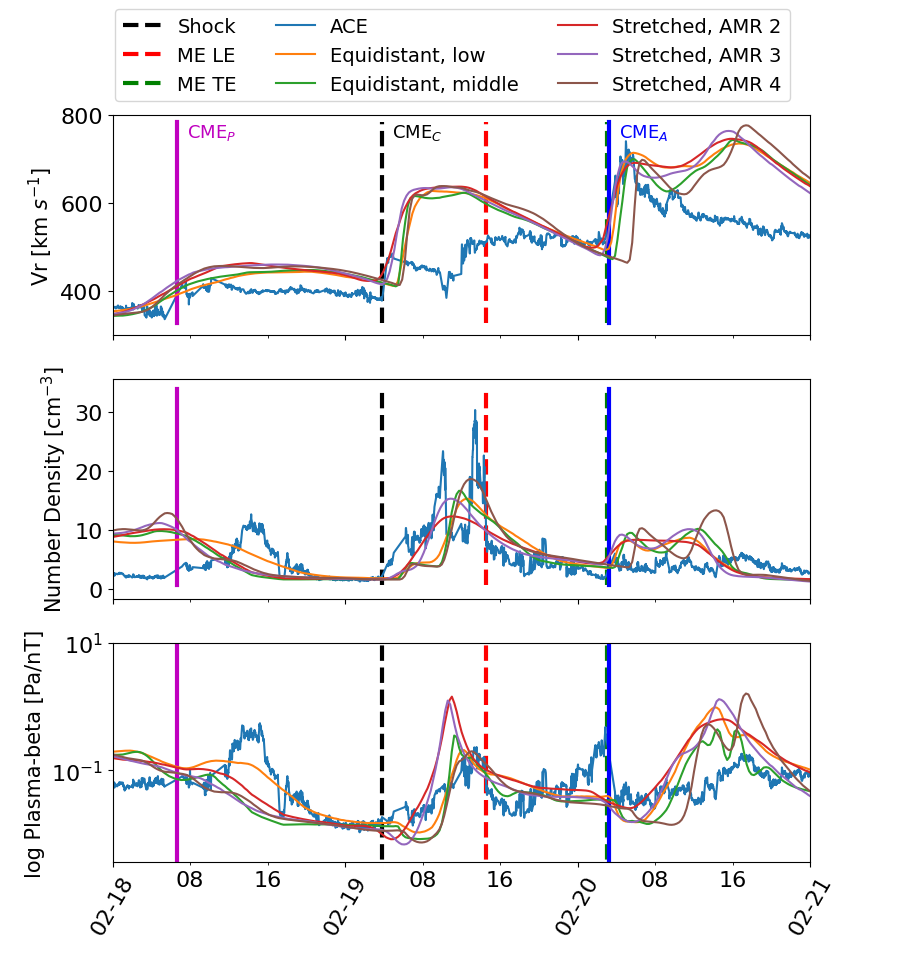}
    \includegraphics[width=0.48\textwidth]{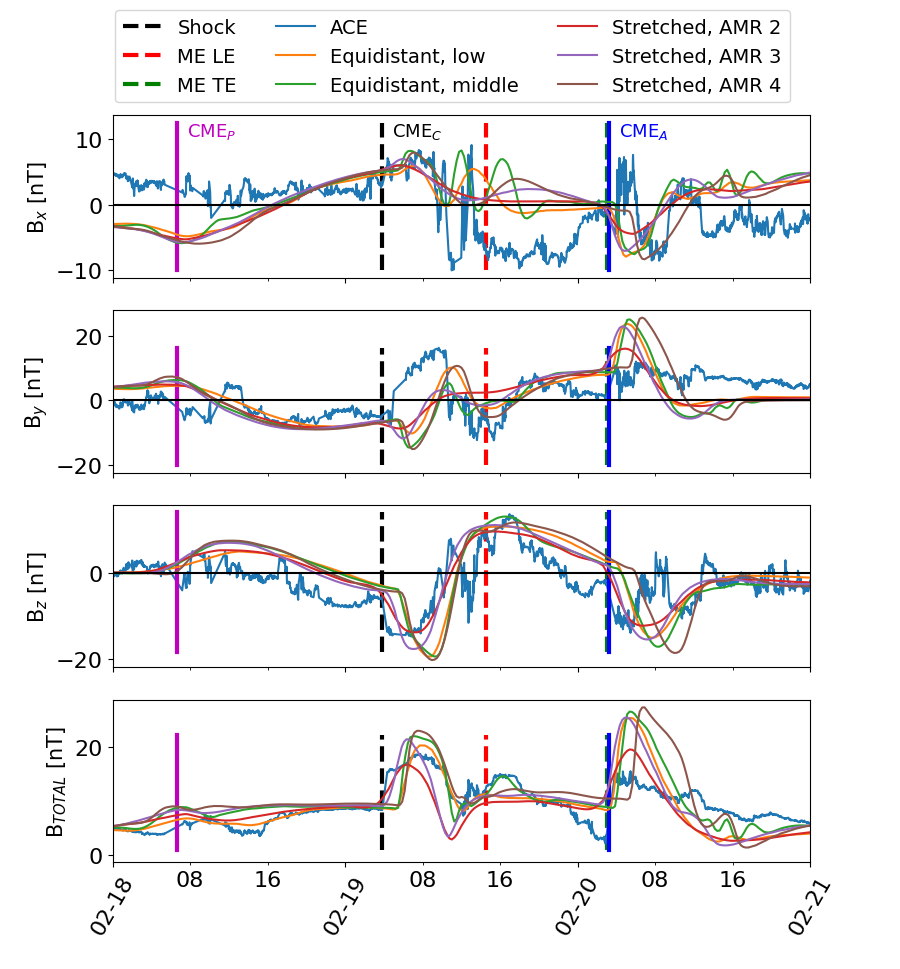}
    \caption{Simulations performed with different numerical grids in Icarus. The blue line corresponds to in-situ data observed at the ACE spacecraft. The radial velocity, number density, and logarithm of plasma-$\beta$ are plotted on the left figure. The right figure shows the magnetic field components and the total magnetic field. The shock arrival time, the magnetic ejecta leading edge (ME LE), and the magnetic ejecta trailing edge (ME TE) are indicated in the catalog from \cite{winslow2015} and are plotted with black, red, and green dashed lines, respectively. The solid vertical magenta and blue lines indicate the arrival of the shock disturbances of the previous and following CMEs on Earth. Equidistant grid simulations on the low and middle-resolution grids are given in orange and green. The simulations performed on the stretched grid with AMR levels 2, 3, and 4 are given in red, purple, and brown, respectively.}
    \label{fig:2014_ace_v_timeseries}
\end{figure*}

The CME arrives later than in the observed data in the simulation results. The time delay is $10-13$ hours, depending on the resolution of the simulation. \cite{Grison2018} describes the deceleration of the CMEs beyond the Mercury orbit until 1~au. In this case, the CME deceleration was significant, as from ballistic measurements of the CME speed, the CME transit speed from the Sun to Mercury was estimated to be $\sim 820\;$km s$^{-1}$, and the average CME transit speed from Mercury to Earth was estimated to be $\sim 570\;$km s$^{-1}$ \citep{Salman2020}. In the simulation, this deceleration was not recovered. Therefore, the arrival times at the two locations were not accurate. Since the observed data was of low quality and mostly missing at Messenger, we fixed the parameters to match the arrival time at Earth. Thus, we get a significant delay in the synthetic data for the CME arrival at MESSENGER. For more accessible comparison purposes on the right panel in Figure~\ref{fig:2014_messenger_b_timeseries}, we shifted the synthetic data by 13 hours, as mentioned. We only plotted the results of one simulation corresponding to AMR level 4 so as not to overload the figure. This figure shows that the polarities of the magnetic field components match the observed data. The negative values of the $B_z$ component are not reconstructed well, and the strength of the positive values is similar to the observed data. 

The left panel of Figure~\ref{fig:2014_messenger_b_timeseries} displays results computed on different simulation grids. The low resolution and AMR level 2 simulations show similar smooth profiles. The AMR level 3 simulation already indicates results comparable to the equidistant medium-resolution simulation. The AMR level 4 simulation result demonstrates the sharpest profiles when changing the magnetic field polarities, which can also be seen in the total magnetic field strength profile. Unfortunately, further comparison of the synthetic data with the observations at MESSENGER is difficult, as most of the observed data is missing. Therefore, the passing of the ME is ambiguous since it coincides with the period when the MESSENGER is located within the planet's magnetosphere.

Figure~\ref{fig:2014_ace_v_timeseries} shows the time profiles at ACE. The observational data by ACE are plotted in blue, and the vertical dashed black, red, and green lines represent the arrival of the shock disturbance, ME LE and ME TE at Earth, respectively, for CME$_C$. The arrival times for the shock and the passing of the ME are the same as indicated in \cite{winslow2015}. However, the arrival of the ME was shifted from the reported 12:00 to 14:34, as the plasma variables in these $\sim2.5$ hours have more of a sheath structure than of the ME. The plasma-$\beta$ and magnetic field components fluctuate, usually characteristic of the sheath region. After the newly identified arrival of the ME, the ME has more homogeneous profiles, and plasma-$\beta$ values start to decrease, indicating the domination of the magnetic forces over the plasma pressure force. 

In Figure~\ref{fig:2014_ace_v_timeseries}, the solid vertical magenta and blue lines indicate the arrival of the sheath of the previous and following CMEs, indicated with CME$_P$ and CME$_A$. 
Since the shock of CME$_A$ arrives shortly after the passage of the ME TE of the CME$_C$, they appear very close on the figure, but the lines can be distinguished.  

At the magenta vertical line, corresponding to the arrival of CME$_P$ in the simulation, we can see a slight increase in the speed profile, comparable to that in the observational data. The enhancement in the number density profiles is also localized around this vertical magenta line. In contrast, in the observational data, the main increase in the number density arrives later at ACE. The plasma-$\beta$ values indicate the arrival of the ME shortly after the arrival of the magneto sheath region, which is in good agreement with the observed data. The modeled magnetic field components agree with the observations for the first CME. The strength of the total magnetic field is also similar to the one observed, guaranteeing good background conditions upon the arrival of the following CME, which is the main CME we are modeling in this study. The profiles in this region are smooth. Therefore, no significant advantage can be seen in any of the different simulations regarding the sharpness of profiles or resolving more small-scale structures. 

The arrival time for the second CME, indicated by CME$_C$ on the figure, is similar in the synthetic modeled data and the observations. However, the solar wind profile is overestimated before the arrival by $\sim 40\;$km s$^{-1}$. 
The shock jump is calculated by the following formula $\text{V}_\text{jump} = \text{V}_\text{P} - \text{V}_\text{SW}$, where SW stands for the solar wind before the arrival of the disturbance and $P$ stands for the peak value upon the arrival. From this calculation, the speed jump in the observational data is $\sim 105\;$km s$^{-1}$, while in the synthetic data from the simulation, it corresponds to $\sim 210\;$km s$^{-1}$. Therefore, the speed upon the arrival of the CME is strongly overestimated, which indicates that the deceleration is not very well modeled in Icarus. 

When the ME arrives at ACE, there is a further increase in the speed values in the observed data by $\sim60\;$km s$^{-1}$. Then, the profile of the speed is flat while passing by ACE, while in the synthetic data from the simulation, the speed values decrease towards the tail of the CME, obtaining similar values at the ME TE. 
This could indicate that the CME was more compressed in the observed data by the third CME, not allowing it to expand, while the compression was not reconstructed to the same extent in the modeled data. 

The number density modeled by Icarus agrees with the observed data. 
In plasma-$\beta$ values, we can see that the values in the modeled data start to decrease, similar to the observed data, where the arrival of the ME LE is indicated. However, we see that in the observed data, the values start to increase at the ME rear, 
which is not the case in the modeled data. This indicates that the third CME has not compressed the second one as much in the simulation as in the actual scenario. 

\begin{figure*}
    \centering
    \includegraphics[width=0.33\textwidth]{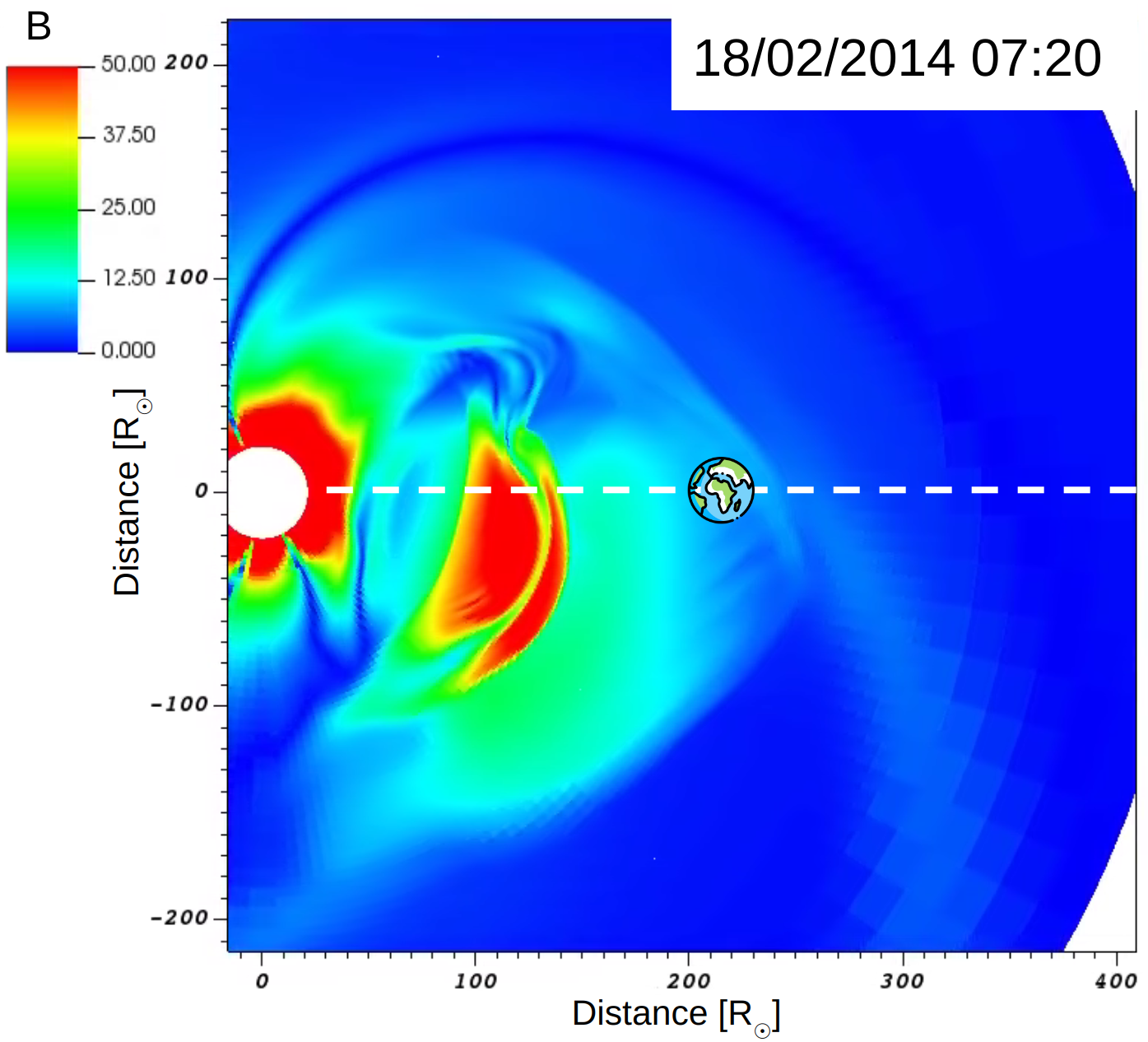}
    \includegraphics[width=0.33\textwidth]{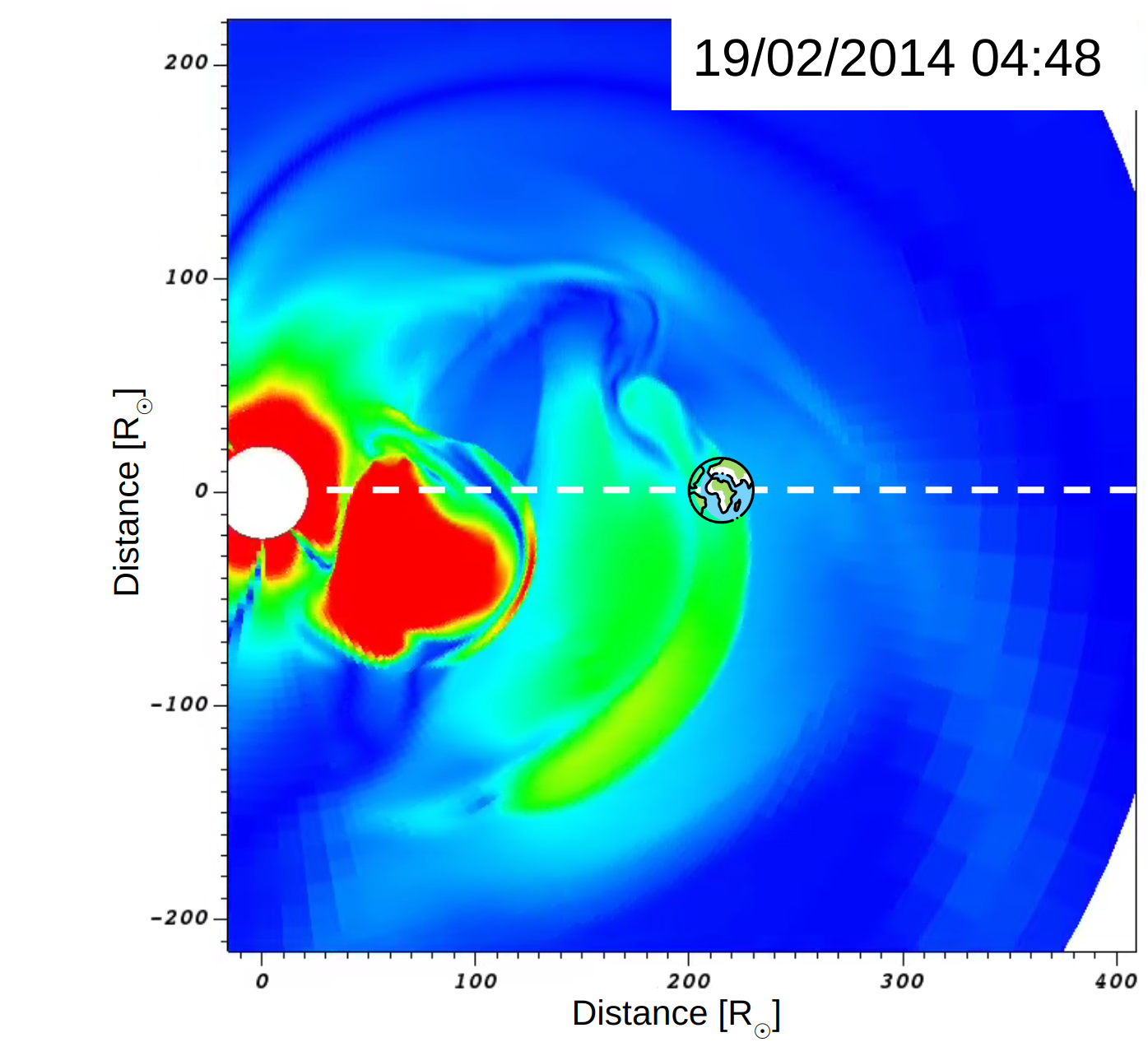}
    \includegraphics[width=0.33\textwidth]{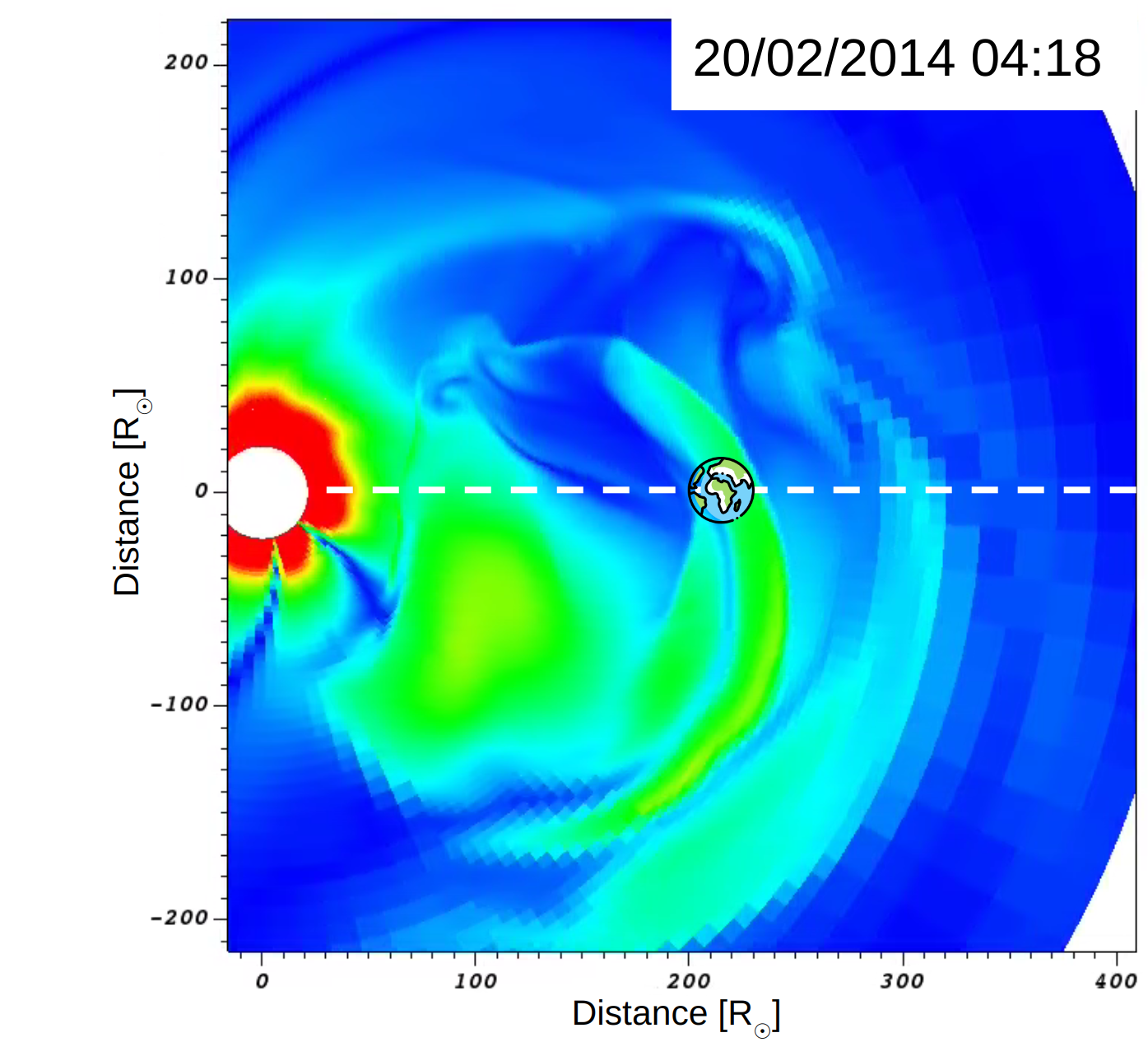}
    \caption{The snapshots represent the equatorial plane in Icarus. The slice is colored with B - total magnetic field values in [nT]. The distance from the Sun is indicated on the horizontal and vertical axes. The left figure represents a snapshot taken at the arrival time of the first CME on Earth. The second CME represents the snapshot corresponding to the arrival of the second CME, and the third one denotes the magnetic field configuration upon the arrival of the third CME. The timestamps are indicated on each snapshot. The frame is rotated so that Earth is fixed at 0 longitude. The white dashed line is along the Sun-Earth line.}
    \label{fig:2014_icarus_b}
\end{figure*}

When considering the total magnetic field strength (Figure~\ref{fig:2014_ace_v_timeseries}, left bottom panel), there is a substantial increase upon the arrival of the shock of the second CME. This results from the CME-CME interaction occurring from the orbit of Mercury onwards to 1~au, as the shock of the second CME travels through the ME of the first one. 
This region is slightly more compressed in the modeled data than in the observations. Still, all magnetic field components are very similar to the observed data, apart from the $B_y$ component, for which the negative region is more expanded, and the positive region seems more compressed and arrives later. The magnetic field values before the arrival of the ME are in agreement with the observations. The ME arrives at the red vertical dashed line in Figure~\ref{fig:2014_ace_v_timeseries}, both in the simulations and observation data. 
The modeled and observed profiles of the magnetic field components are similar. In the simulation results, the profiles seem more ``stretched'', caused by a weaker interaction with the third CME. 

The simulated third CME, CME$_A$, arrives simultaneously as in the observed data. Only the first part of the third CME is plotted, as analyzing the third CME in detail exceeds the scope of this paper. Because of the over-expanded second CME, we obtain a CME-CME interaction between the second and the third CME, which is much smaller in the observed data. The speed values are in good agreement between the observed and modeled data in the front part of CME$_A$; however, later on, the simulated CME is much faster. 
The plasma$-\beta$ values are slightly lower, and the total magnetic field is also stronger in the simulations than in the observations in the front part of CME$_A$, while it is the reverse later on. 

The interactions with the other CMEs were not captured well with a smaller radius for the second CME. The reason for the over-expanded second CME and insufficient compression can be a characteristic of the spheromak CME model, which usually is more radially expanded than in the observations \citep{Scolini2019}. 

The separation between the spacecraft during this CME case was $4.8^\circ$, which is slightly larger than in the first case. When comparing the magnetic field components, the polarities are in good agreement for $B_y$ and $B_z$, but are different for the $B_x$ component, which the longitudinal separation between the spacecraft could cause. 

The refinement criterion was the same as in the first CME case reported in Section~\ref{2013_july}, with only one difference, as we fixed a slightly larger margin for the latitudinal direction to allow more refinement.   Upon the arrival of the second CME, the shock jump in the speed values is the sharpest in the simulation performed on the medium-resolution equidistant grid and in the AMR level 4 simulation (Figure~\ref{fig:2014_ace_v_timeseries}).  
The AMR level 3 results are similar to the AMR level 2 simulations regarding smoothness. The third CME arrives at the latest in the AMR level 4 simulation, but the profile is very similar to that from the medium equidistant simulation. 
Therefore, we can see that the features are well resolved at Mercury with AMR level 3 simulation. In contrast, at Earth, the profiles modeled with AMR level 3 are relatively smooth, and an additional refinement level is required to see smaller structures. 

Figure~\ref{fig:2014_icarus_b} shows the simulation results in the equatorial plane crossing Earth, showing the strength of the magnetic field in nT. 
The three snapshots are taken at the arrival times at 1~au for CME$_P$, CME$_C$, and CME$_A$, respectively. The simulation output is saved only every 3 hours. Therefore, the snapshot closest to the arrival times is plotted. All three snapshots are within 1 hour from the observed arrival time reported in Table~\ref{table:3cmes}. The simulation is performed on an AMR level 4 grid.
This figure shows the increase in the total magnetic field regions upon the arrival of the second and third CMEs. In the middle and right panels, we can see the CME-CME interactions between the first and second CME pairs and the second and third CME pairs, respectively. 


\begin{table}[t!]
\caption{The wall-clock times for the performed simulations in Icarus.}   
\label{table:2014_run_times}   
\centering            
\begin{tabular}{c c c c c}         
\hline
Low$_\text{EQ}$& Middle$_\text{EQ}$ & AMR 2 & AMR 3 & AMR 4  \\ 
\hline\hline
  7m 9s & 1h 12m 42s & 4m 11s & 9m 0s & 38m 39s  \\ \hline                                 
\end{tabular}
\tablefoot{
 The runs are performed with different computational grids on four nodes with 2 Xeon Gold 6240 CPUs@2.6 GHz (Cascadelake), 18 cores each, on the Genius cluster at KU Leuven. Low$_\text{EQ}$ and Middle$_\text{EQ}$ correspond to simulations on the equidistant low and medium-resolution grids. 
}
\end{table}

Table~\ref{table:2014_run_times} shows the run times for the simulations performed in Icarus. All simulations are performed on four nodes on the Genius cluster at KU Leuven. The medium-resolution simulation is 8 and 1.9 times slower than the simulation with the stretched grid and AMR levels 3 and 4. AMR level 3 produces results similar to the medium equidistant simulation at Mercury, and AMR level 4 performs similarly to the medium-resolution simulation on Earth. Therefore, depending on the purpose of the simulation, the maximum refinement level can be chosen accordingly. 

\subsection{Numerical scaling tests}

\begin{figure}[h!]
    \centering
    \includegraphics[width=0.5\textwidth]{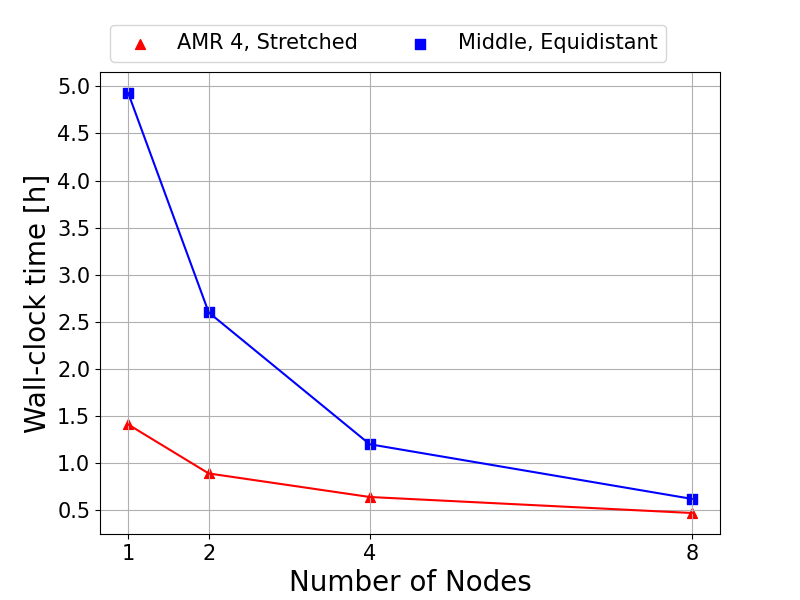}
    \caption{The scaling tests for the medium-resolution simulation on the equidistant grid and the AMR level 4 simulation on the stretched grid. The horizontal axis denotes the number of nodes, and the vertical axis denotes the wall clock time in hours. The simulations are performed on the given amount of nodes with 2 Xeon Gold 6240 CPUs@2.6 GHz (Cascadelake), 18 cores each, on the Genius cluster at KU Leuven.}
    \label{fig:2014_scaling_test}
\end{figure}

Figure~\ref{fig:2014_scaling_test} compares the time performance of the two types of simulations. The highest resolution simulations, the AMR level 4 run on a stretched grid and the medium-resolution simulation on an equidistant grid, were performed on 1, 2, 4, and 8 nodes with 2 Xeon Gold 6240 CPUs@2.6 GHz (Cascadelake), 18 cores each, on the Genius cluster at KU Leuven. The AMR level 4 and medium-resolution simulation times are marked with red triangles and blue squares, respectively. When both simulations are performed on one node, AMR level 4 simulation is 3.5 times faster; on two nodes - 2.9 times faster; on four nodes - 1.88 times faster; and on eight nodes - 1.3 times faster. 

Additionally, we calculated that performing the medium-resolution simulation on four nodes speeds up simulations by a factor of 4.1 compared to the performed simulation on one node. In contrast, for the AMR level 4 case, the simulation on four nodes is 2.2 times faster than on one node. The medium and AMR level 4 simulations performed on eight nodes are 7.95 and 3 times faster than those performed on one node. Therefore, the efficiency gain when running with additional nodes decreases with more AMR levels than in the medium-resolution equidistant simulation. Nevertheless, the simulations performed on the stretched grid with AMR level 4 are faster than those with the medium-resolution simulation on the equidistant grid while resolving similarly or better the small-scale features in the variations of the time series of the modeled quantities.

\section{Conclusions} \label{conclusions}

The newly implemented magnetized linear force-free spheromak model was validated inside the optimized heliospheric wind and CME evolution tool Icarus. To do so, we selected two CME events that were observed by two spacecraft, namely MESSENGER and ACE. 

There were multiple reasons to choose these particular events. First, the CME events observed by MESSENGER and ACE were favorable, since the radial separation between the orbit of Mercury and Earth is large as it ranges between $0.55 - 0.65$~au, depending on the location of the planets. This allows a comparison of the performance of the code close to the inner boundary of the computational domain and farther out in the middle of the domain (in the radial direction). 

The events were chosen from the catalog described in \citet{winslow2015} and had already been further analyzed by \citet{Grison2018}. From the listed events, we chose one that represents a quiet case, with a clearly defined CME shock and magnetic cloud, and a second one where more complex interactions are observed. The first event occurred on July 9, 2013, and the second on February 16, 2014. 

We identified the source region for the first CME event and obtained the parameters for modeling the CME with the geometrical fitting on the StereoCAT online tool and 3D GCS reconstruction. Then we injected the spheromak CME model in Icarus with the determined parameters. The synthetic (modeled) data at the locations of Mercury and Earth were compared to the observed data by MESSENGER and ACE. The periods MESSENGER spent in Mercury's magnetosphere were removed from the time series. However, the arrival time of the CME at MESSENGER coincided with the spacecraft passing through the planet's magnetosphere. Therefore, it can not be considered the start of the magnetosheath of the CME at MESSENGER. The spacecraft goes through the magnetosphere of Mercury two more times until the passing of the trailing edge is registered at MESSENGER, therefore the data are not complete. 
Still, Icarus modeled the polarities of the magnetic field components in agreement with the observed data. Unfortunately, MESSENGER does not observe plasma conditions like speed, number density, and temperature, and hence, complete validation of all states was impossible. 

Further, we investigated the effect of the CME at ACE. We modeled the arrival time of the CME in agreement with the observed data, and the modeled magnetic field components also replicated the strength values and polarity signs of the observed data. As reported by \citet{Grison2018}, no significant deceleration/acceleration was obtained for this CME event. The arrival at both locations agreed with the dates reported in \cite{winslow2015} and \cite{Salman2020}. The simulation performed using the AMR level 4 grid resolved the small-scale signatures on Earth, while at Mercury, the profiles were rather smooth, and no significant advantage of using AMR was spotted. The AMR level 3 and 4 simulations were, respectively, $\sim9.5$ and $2.1$ times faster than the medium-resolution equidistant grid simulation, even though the results modeled with the AMR level 3  and 4 simulations were similar or slightly better than the medium-resolution simulations on the equidistant grid.

The second CME case was more complex, as the catalog by \citet{richardson_catalog} registered other CMEs before and after the chosen CME. We identified the source of the second CME event and obtained the CME parameters again through StereoCAT and GCS fitting. The coronagraph images did not provide sufficiently high-quality data, and the CME's contributions were faint. Therefore, the CME parameters were further optimized with heliospheric modeling. However, it was impossible to reconstruct the time series profiles by modeling the isolated CME in Icarus. 

The conditions prior to the arrival of the CME were significantly different in the simulations, and the CME was overexpanded. 
Therefore, the preceding and following CMEs were taken into account in the Icarus model. The source of the preceding CME could not be identified, as the CME was slow and extended. 
We estimated the parameters for the injection at 0.1~au, the inner heliospheric boundary, to reconstruct the conditions before the arrival of the CME of our choice. Next, the source region for the third CME was located. A large filament was observed with multiple ribbons, but the identification of the exact source for the ejected flux rope was impossible. The tilt of the spheromak was further optimized with the heliospheric simulations. Since the aim of including the third CME was to investigate its effect on the propagation of the main CME in the event, the parameters of the third CME were not optimized in detail. 

In summary, we modeled the second CME event by injecting three CMEs into Icarus. The CME-CME interaction between the preceding and the main CME was well reconstructed and the arrival signatures were captured in the heliospheric simulation. The main CME was more expanded in the Icarus simulation than in the observations. Therefore, the interaction between the second and third CMEs was underestimated in the simulation results compared to observations. However, we could see that the presence of the third CME still compressed and pushed the second CME, compared to simulations where the third CME was not modeled. The choice of CME density for injecting the CMEs was also important, as it had a large effect on whether the plasma cloud was contained or not. 

Overall, we could notice that less AMR levels are sufficient at the orbit of MESSENGER to distinguish smaller scale features, while more AMR levels are needed at ACE. This is due to the nature of the stretched grid, as the cell sizes are naturally smaller at the orbit of Mercury than at the orbit of Earth. The best results were obtained with AMR level 4 at ACE, while at Mercury AMR level 3 results were already similar to equidistant medium-resolution simulations. This implies that when choosing recent missions for sampling the heliosphere, less AMR levels are sufficient to obtain sufficiently detailed time series when the spacecraft is close to the inner heliospheric boundary. 

We also analyzed the simulation wall-clock times. The AMR level 3 and 4 simulations were 8 and 1.88 times faster than the medium-resolution equidistant simulation runs, respectively. The AMR simulations were slightly slower than in the previous CME case, as three CME were modeled, and a considerably larger part of the computational domain needed to be refined. 

Finally, we performed scaling tests for the AMR level 4 and medium-resolution simulations. For this purpose, we performed simulations on 1, 2, 4, and 8 nodes on the same cluster at KU Leuven. The results show that the scaling is better in the equidistant medium-resolution simulation, as the presence of different AMR levels slows down the parallel computations. 

In summary, the two CME events were modeled successfully in the new heliospheric modeling tool Icarus. The time series profiles were compared to the observed data for both cases. The isolated CME and the complex three-CME interactions were sampled at two locations inside the heliosphere. The results showed that the observed deceleration was not reproduced by the Icarus simulations for the CME-CME interaction case. However, the profiles of the variables in the time series matched the observed data. Future work will focus on studying and modeling the possible deceleration mechanisms in the heliosphere, and a more advanced magnetized CME model, with a self-similar expansion, will be considered for the multi-spacecraft study.

\begin{acknowledgements}
This research has received funding from the European Union’s Horizon 2020 research and innovation programme under grant agreement No 870405 (EUHFORIA 2.0) and the ESA project "Heliospheric modelling techniques“ (Contract No. 4000133080/20/NL/CRS).
These results were also obtained in the framework of the projects C14/19/089  (C1 project Internal Funds KU Leuven), G0B5823N and G002523N  (FWO-Vlaanderen), 4000134474 (SIDC Data Exploitation, ESA Prodex-12), and Belspo project B2/191/P1/SWiM.
The Computational resources and services used in this work were provided by the VSC-Flemish Supercomputer Center, funded by the Research Foundation Flanders (FWO) and the Flemish Government-Department EWI. We acknowledge the Community Coordinated Modeling Center (CCMC) at Goddard Space Flight Center for the use of the StereoCAT tool, \url{https://ccmc.gsfc.nasa.gov/analysis/stereo/}. 
\end{acknowledgements}

%
%

\bibliographystyle{aa}
\bibliography{bibliography}

\begin{thebibliography}{44}
\expandafter\ifx\csname natexlab\endcsname\relax\def\natexlab#1{#1}\fi

\bibitem[{{Arge} {et~al.}(2003){Arge}, {Odstrcil}, {Pizzo}, \&
  {Mayer}}]{arge2003}
{Arge}, C.~N., {Odstrcil}, D., {Pizzo}, V.~J., \& {Mayer}, L.~R. 2003, in
  American Institute of Physics Conference Series, Vol. 679, Solar Wind Ten,
  ed. M.~{Velli}, R.~{Bruno}, F.~{Malara}, \& B.~{Bucci}, 190--193

\bibitem[{{Asvestari} {et~al.}(2022){Asvestari}, {Rindlisbacher}, {Pomoell}, \&
  {Kilpua}}]{Asvestari2022}
{Asvestari}, E., {Rindlisbacher}, T., {Pomoell}, J., \& {Kilpua}, E. K.~J.
  2022, \apj, 926, 87

\bibitem[{{Baratashvili} \& {Poedts}(2024)}]{Baratashvili2024}
{Baratashvili}, T. \& {Poedts}, S. 2024, arXiv e-prints, arXiv:2401.02504

\bibitem[{{Baratashvili} {et~al.}(2022{\natexlab{a}}){Baratashvili}, {Verbeke},
  {Keppens}, \& {Poedts}}]{Baratashvili2022sungeo}
{Baratashvili}, T., {Verbeke}, C., {Keppens}, R., \& {Poedts}, S.
  2022{\natexlab{a}}, Sun and Geosphere, 17

\bibitem[{{Baratashvili} {et~al.}(2022{\natexlab{b}}){Baratashvili}, {Verbeke},
  {Wijsen}, \& {Poedts}}]{Baratashvili2022}
{Baratashvili}, T., {Verbeke}, C., {Wijsen}, N., \& {Poedts}, S.
  2022{\natexlab{b}}, \aap, 667, A133

\bibitem[{{Brueckner} {et~al.}(1995){Brueckner}, {Howard}, {Koomen},
  {Korendyke}, {Michels}, {Moses}, {Socker}, {Dere}, {Lamy}, {Llebaria},
  {Bout}, {Schwenn}, {Simnett}, {Bedford}, \& {Eyles}}]{Brueckner1995}
{Brueckner}, G.~E., {Howard}, R.~A., {Koomen}, M.~J., {et~al.} 1995, \solphys,
  162, 357

\bibitem[{{Burlaga} {et~al.}(1982){Burlaga}, {Klein}, {Sheeley}, {Michels},
  {Howard}, {Koomen}, {Schwenn}, \& {Rosenbauer}}]{Bourlaga1982}
{Burlaga}, L.~F., {Klein}, L., {Sheeley}, N.~R., J., {et~al.} 1982, \grl, 9,
  1317

\bibitem[{{Burlaga}(1991)}]{Bourlaga1991}
{Burlaga}, L. F.~E. 1991, in Physics of the Inner Heliosphere II, Vol.~21,
  1--22

\bibitem[{{Cane} \& {Richardson}(2003)}]{Cane2003}
{Cane}, H.~V. \& {Richardson}, I.~G. 2003, Journal of Geophysical Research
  (Space Physics), 108, 1156

\bibitem[{{Chandrasekhar}(1956)}]{Chandrasekhar1956}
{Chandrasekhar}, S. 1956, Proceedings of the National Academy of Science, 42, 1

\bibitem[{{Davies} {et~al.}(2020){Davies}, {Forsyth}, {Good}, \&
  {Kilpua}}]{Davies2020}
{Davies}, E.~E., {Forsyth}, R.~J., {Good}, S.~W., \& {Kilpua}, E. K.~J. 2020,
  \solphys, 295, 157

\bibitem[{{D{\'e}moulin} \& {Pariat}(2009)}]{demoulin2009}
{D{\'e}moulin}, P. \& {Pariat}, E. 2009, Advances in Space Research, 43, 1013

\bibitem[{Gieseler {et~al.}(2023)Gieseler, Dresing, Palmroos, Freiherr~von
  Forstner, Price, Vainio, Kouloumvakos, Rodríguez-García, Trotta, Génot,
  Masson, Roth, \& Veronig}]{solarmach}
Gieseler, J., Dresing, N., Palmroos, C., {et~al.} 2023, Frontiers in Astronomy
  and Space Sciences, 9

\bibitem[{{Gopalswamy} {et~al.}(2017){Gopalswamy}, {Yashiro}, {Akiyama}, \&
  {Xie}}]{Gopalswamy2017}
{Gopalswamy}, N., {Yashiro}, S., {Akiyama}, S., \& {Xie}, H. 2017, \solphys,
  292, 65

\bibitem[{{Grison} {et~al.}(2018){Grison}, {Sou{\v{c}}ek}, {Krupar},
  {P{\'\i}{\v{s}}a}, {Santol{\'\i}k}, {Taubenschuss}, \&
  {N{\u{e}}mec}}]{Grison2018}
{Grison}, B., {Sou{\v{c}}ek}, J., {Krupar}, V., {et~al.} 2018, Journal of Space
  Weather and Space Climate, 8, A54

\bibitem[{{James} {et~al.}(2017){James}, {Imber}, {Bunce}, {Yeoman},
  {Lockwood}, {Owens}, \& {Slavin}}]{james2017}
{James}, M.~K., {Imber}, S.~M., {Bunce}, E.~J., {et~al.} 2017, Journal of
  Geophysical Research (Space Physics), 122, 7907

\bibitem[{{Kaiser}(2005)}]{Kaiser2005}
{Kaiser}, M.~L. 2005, Advances in Space Research, 36, 1483

\bibitem[{{Kilpua} {et~al.}(2017){Kilpua}, {Koskinen}, \&
  {Pulkkinen}}]{Kilpua2017}
{Kilpua}, E., {Koskinen}, H. E.~J., \& {Pulkkinen}, T.~I. 2017, Living Reviews
  in Solar Physics, 14, 5

\bibitem[{{Maharana} {et~al.}(2023){Maharana}, {Scolini}, {Schmieder}, \&
  {Poedts}}]{Maharana2023}
{Maharana}, A., {Scolini}, C., {Schmieder}, B., \& {Poedts}, S. 2023, \aap,
  675, A136

\bibitem[{{Marubashi} {et~al.}(2015){Marubashi}, {Akiyama}, {Yashiro},
  {Gopalswamy}, {Cho}, \& {Park}}]{Marubashi2015}
{Marubashi}, K., {Akiyama}, S., {Yashiro}, S., {et~al.} 2015, \solphys, 290,
  1371

\bibitem[{{National Research Council,}(2008)}]{press_reference}
{National Research Council,}. 2008, Severe Space Weather Events: Understanding
  Societal and Economic Impacts: A Workshop Report (Washington, DC: The
  National Academies Press)

\bibitem[{{Palmerio} {et~al.}(2018){Palmerio}, {Kilpua}, {M{\"o}stl},
  {Bothmer}, {James}, {Green}, {Isavnin}, {Davies}, \&
  {Harrison}}]{Palmerio2018}
{Palmerio}, E., {Kilpua}, E.~K.~J., {M{\"o}stl}, C., {et~al.} 2018, Space
  Weather, 16, 442

\bibitem[{{Palmerio} {et~al.}(2021){Palmerio}, {Nieves-Chinchilla}, {Kilpua},
  {Barnes}, {Zhukov}, {Jian}, {Witasse}, {Provan}, {Tao}, {Lamy}, {Bradley},
  {Mays}, {M{\"o}stl}, {Roussos}, {Futaana}, {Masters}, \&
  {S{\'a}nchez-Cano}}]{Palmerio2021}
{Palmerio}, E., {Nieves-Chinchilla}, T., {Kilpua}, E. K.~J., {et~al.} 2021,
  Journal of Geophysical Research (Space Physics), 126, e2021JA029770

\bibitem[{{Pomoell} \& {Poedts}(2018)}]{Pomoell2018}
{Pomoell}, J. \& {Poedts}, S. 2018, Journal of Space Weather and Space Climate,
  8, A35

\bibitem[{Richardson \& Cane(2024)}]{richardson_catalog}
Richardson, I. \& Cane, H. 2024

\bibitem[{{Salman} {et~al.}(2020){Salman}, {Winslow}, \& {Lugaz}}]{Salman2020}
{Salman}, T.~M., {Winslow}, R.~M., \& {Lugaz}, N. 2020, Journal of Geophysical
  Research (Space Physics), 125, e27084

\bibitem[{{Sarkar} {et~al.}(2024){Sarkar}, {Pomoell}, {Kilpua}, {Asvestari},
  {Wijsen}, {Maharana}, \& {Poedts}}]{Sarkar2024}
{Sarkar}, R., {Pomoell}, J., {Kilpua}, E., {et~al.} 2024, \apjs, 270, 18

\bibitem[{{Scolini} {et~al.}(2019){Scolini}, {Rodriguez}, {Mierla}, {Pomoell},
  \& {Poedts}}]{Scolini2019}
{Scolini}, C., {Rodriguez}, L., {Mierla}, M., {Pomoell}, J., \& {Poedts}, S.
  2019, \aap, 626, A122

\bibitem[{{Sun} {et~al.}(2022){Sun}, {Dewey}, {Aizawa}, {Huang}, {Slavin},
  {Fu}, {Wei}, \& {Bowers}}]{sun2022}
{Sun}, W., {Dewey}, R.~M., {Aizawa}, S., {et~al.} 2022, Science China Earth
  Sciences, 65, 25

\bibitem[{{Thernisien}(2011)}]{Thernisien2011}
{Thernisien}, A. 2011, \apjs, 194, 33

\bibitem[{{THIRA}(2019)}]{THIRA2019}
{THIRA}. 2019, 2019 National Threat and Hazard Identification and Risk
  Assessment (THIRA)

\bibitem[{{Titov} \& {D{\'e}moulin}(1999)}]{titov1999}
{Titov}, V.~S. \& {D{\'e}moulin}, P. 1999, \aap, 351, 707

\bibitem[{{T{\'o}th} \& {Odstr{\v{c}}il}(1996)}]{Toth1996}
{T{\'o}th}, G. \& {Odstr{\v{c}}il}, D. 1996, Journal of Computational Physics,
  128, 82

\bibitem[{{van Leer}(1977)}]{vanleer1977}
{van Leer}, B. 1977, Journal of Computational Physics, 23, 276

\bibitem[{{Verbeke} {et~al.}(2022){Verbeke}, {Baratashvili}, \&
  {Poedts}}]{Verbeke2022}
{Verbeke}, C., {Baratashvili}, T., \& {Poedts}, S. 2022, \aap, 662, A50

\bibitem[{{Verbeke} {et~al.}(2019){Verbeke}, {Pomoell}, \&
  {Poedts}}]{Verbeke2019}
{Verbeke}, C., {Pomoell}, J., \& {Poedts}, S. 2019, \aap, 627, A111

\bibitem[{{Vourlidas} {et~al.}(2013){Vourlidas}, {Lynch}, {Howard}, \&
  {Li}}]{Vourlidas2013}
{Vourlidas}, A., {Lynch}, B.~J., {Howard}, R.~A., \& {Li}, Y. 2013, \solphys,
  284, 179

\bibitem[{{Wang} {et~al.}(2018){Wang}, {Shen}, {Liu}, {Liu}, {Guo}, {Li}, {Xu},
  {Hu}, \& {Zhang}}]{Wang2018}
{Wang}, Y., {Shen}, C., {Liu}, R., {et~al.} 2018, Journal of Geophysical
  Research (Space Physics), 123, 3238

\bibitem[{{Wang} \& {Sheeley}(1990)}]{wang1990}
{Wang}, Y.~M. \& {Sheeley}, N.~R., J. 1990, \apj, 355, 726

\bibitem[{{Webb} \& {Howard}(2012)}]{Webb2012}
{Webb}, D.~F. \& {Howard}, T.~A. 2012, Living Reviews in Solar Physics, 9, 3

\bibitem[{Webb {et~al.}(2006)Webb, Mizuno, Buffington, Cooke, Eyles, Fry,
  Gentile, Hick, Holladay, Howard, Hewitt, Jackson, Johnston, Kuchar, Mozer,
  Price, Radick, Simnett, \& Tappin}]{Webb2006}
Webb, D.~F., Mizuno, D.~R., Buffington, A., {et~al.} 2006, Journal of
  Geophysical Research: Space Physics, 111

\bibitem[{{Winslow} {et~al.}(2015){Winslow}, {Lugaz}, {Philpott}, {Schwadron},
  {Farrugia}, {Anderson}, \& {Smith}}]{winslow2015}
{Winslow}, R.~M., {Lugaz}, N., {Philpott}, L.~C., {et~al.} 2015, Journal of
  Geophysical Research (Space Physics), 120, 6101

\bibitem[{{Xia} {et~al.}(2018){Xia}, {Teunissen}, {El Mellah}, {Chan{\'e}}, \&
  {Keppens}}]{Xia2018}
{Xia}, C., {Teunissen}, J., {El Mellah}, I., {Chan{\'e}}, E., \& {Keppens}, R.
  2018, \apjs, 234, 30

\bibitem[{{Zhang} {et~al.}(2021){Zhang}, {Temmer}, {Gopalswamy}, {Malandraki},
  {Nitta}, {Patsourakos}, {Shen}, {Vr{\v{s}}nak}, {Wang}, {Webb}, {Desai},
  {Dissauer}, {Dresing}, {Dumbovi{\'c}}, {Feng}, {Heinemann}, {Laurenza},
  {Lugaz}, \& {Zhuang}}]{Zhang2021}
{Zhang}, J., {Temmer}, M., {Gopalswamy}, N., {et~al.} 2021, Progress in Earth
  and Planetary Science, 8, 56

\end{thebibliography}

\end{document}